\documentclass[useAMS,usenatbib]{mn2e}

\usepackage{epsfig}
\usepackage{amsmath}
\usepackage{amssymb}

\newcommand{\sF}{\mathcal{F}}

\newcommand{\sS}{\mathcal{S}}
\newcommand{\sT}{\mathcal{T}}

\newcommand{\bA}{\mbox{\boldmath$A$}}

\newcommand{\bF}{\mbox{\boldmath$F$}}
\newcommand{\bX}{\mbox{\boldmath$X$}}
\newcommand{\be}{\mbox{\boldmath$e$}}
\newcommand{\bg}{\mbox{\boldmath$g$}}

\newcommand{\br}{\mbox{\boldmath$r$}}

\newcommand{\bu}{\mbox{\boldmath$u$}}

\newcommand{\dd}{{\rm d}}
\newcommand{\half}{{\textstyle\frac{1}{2}}}

\def\pre{Phys.~Rev.~Lett.}
\def\aap{A\&A}
\def\apj{ApJ}
\def\mnras{MNRAS}
\def\araa{ARA\&A}

\title[Turbulent convection]
{A model of the entropy flux and Reynolds stress in turbulent convection}

\author[P. Garaud, G. I. Ogilvie, N. Miller, S. Stellmach]
  {P. Garaud$^{1}$, G. I. Ogilvie$^{2}$, N. Miller$^{3}$, and 
   S. Stellmach$^{1,4}$ \\
  $^1$Department of Applied Mathematics and Statistics, Baskin 
  School of Engineering, University of California, \\
  1156 High Street, Santa Cruz, CA 95064, USA\\
  $^2$Department of Applied Mathematics and Theoretical Physics,
  University of Cambridge, Centre for Mathematical Sciences,\\ Wilberforce Road, Cambridge CB3 0WA \\
  $^3$Department of Astronomy and Astrophysics, University of California, 1156 High Street, Santa Cruz, CA 95064, USA, \\
  $^4$Institut f\"ur Geophysik, Westf\"alische Wilhelms-Universit\"at M\"unster, D-48149 M\"unster, Germany}

\begin{document}

\maketitle

\label{firstpage}

\begin{abstract}
We propose a closure model for the transport of entropy and momentum
in astrophysical turbulence, intended for application to rotating stellar convective
regions. Our closure model
is first presented in the Boussinesq formalism, and compared 
with laboratory and numerical experimental results on Rayleigh-B\'enard
convection and Homogeneous Rayleigh-B\'enard convection. 
The predicted angular momentum transport properties of the turbulence 
in the slowly rotating case recover the well-known $\Lambda-$effect, with
an amplitude uniquely related to the convective heat flux. 
The model is then extended to the anelastic 
case as well as the fully compressible case. In the special case of spherical
symmetry, the predicted radial heat flux is equivalent to that of 
mixing-length theory. For rotating stars, our model describes the coupled
transport of heat and angular momentum, and provides a unified 
formalism in which to study both differential rotation and 
thermal inhomogeneities in stellar convection zones.
\end{abstract}
\begin{keywords}
convection; hydrodynamics; turbulence; stars:rotation
\end{keywords}

\section{Introduction}

Turbulent convection occurs frequently in stellar interiors and other
astrophysical fluid flows.  While convective motion naturally
transports heat and chemical elements, the transport of angular
momentum by convection in rotating bodies is a more subtle issue.  It
is of particular interest in the case of the Sun, where the internal
pattern of rotation has been measured but remains incompletely
understood.  It may also play an significant role in accretion flows.

Numerical simulations of astrophysical convection are becoming
increasingly powerful and capable of resolving a widening range of
length and time scales.  Nevertheless, a simpler, statistical
description of turbulent transport is desirable in order to treat the
effects of convection on the structure and evolution of stars.  It
almost goes without saying that such a description cannot be derived
strictly from the equations of fluid dynamics but must involve some
modelling or parametrization.

The mixing-length theory of turbulent transport was developed by
Prandtl (1925) and applied to stellar convection by Biermann (1932).
It is still the basic model used in most calculations of stellar
structure and evolution, usually in the form devised by B\"ohm-Vitense
(1958).  The main purpose of mixing-length theory is to relate the
convective heat flux to the superadiabatic gradient; in this context
it does not usually deal with the transport of (angular) momentum that
arises in the presence of shear or rotation.

A standard theoretical approach to convection in differentially
rotating stars is set out in the monograph by R\"udiger (1989).
Angular momentum transport is described by a Reynolds stress tensor
whose components can be related to the large-scale mean flows and 
thermodynamical gradients. A first contribution to the Reynolds stress 
is typically proportional to the angular velocity gradient through
a turbulent viscosity coefficient. An important additional contribution
comes from the $\Lambda$-effect (named after Lebedinsky),
whereby even uniformly rotating convection transports angular momentum
by virtue of its anisotropy. Attempts to constrain or parameterize
these quantities have been made through local 
numerical simulations (e.g.\ K\"apyl\"a, Korpi \& Tuominen
2004) or theoretical models (e.g.\ Kitchatinov \& R\"udiger
1993). Mean-field models of stellar rotation
(e.g. Kitchatinov \& R\"udiger 1999, Rempel 2005) have
been developed which use such parameterized expressions for the Reynolds
stress and heat flux.

Reynolds-stress models of turbulent flows have been developed in the
engineering community over several decades (e.g.\ Pope 2000).
The exact equation governing the Reynolds stress in a turbulent fluid
cannot be solved because of the well known closure problem whereby an
infinite hierarchy of correlations is involved.  Nevertheless, by
parametrizing the difficult terms in this equation, models can be
constructed that bear some fidelity to the turbulent dynamics.  From a
more physical point of view, what is obtained is a time-dependent
constitutive equation for the turbulent fluid, which relates the
turbulent stress to the local history of deformation. There is a
close similarity with models of non-Newtonian fluids (Ogilvie \& Proctor 2003). 
The advection and deformation of the turbulent stress are accurately represented
since they derive from linear terms in the Reynolds-stress equation,
while the nonlinear `relaxation' effects are only modelled (as is also
true for non-Newtonian fluids).

A similar approach can be applied to turbulent convection in which
buoyancy forces play an essential role.  The additional correlations
that must be considered are the flux and the variance of entropy (or
temperature, in the Boussinesq approximation).  This approach offers
some benefits over the conventional description in terms of a
turbulent viscosity and a $\Lambda$-effect.  It can be formulated in a
covariant manner and is not tied to the spherical geometry of a slowly
rotating star.  It starts from a more fundamental description and
allows phenomena such as the $\Lambda-$effect to emerge in a natural way
from more elementary considerations.  It may also allow a more unified
approach to be taken towards problems involving astrophysical
turbulence.

In this paper we explore some of the consequences of a simple
dynamical model of astrophysical convection of this type.  The model
derives from one originally conceived for magnetohydrodynamic
turbulence in accretion discs (Ogilvie 2003) and later applied to
rotating shear flows without magnetic fields (Garaud \& Ogilvie 2005,
GO05 herafter).
Our motivation is to develop and test a model that can be applied to
the convective zone of the Sun, to other stars or to accretion discs.
We emphasize, however, that our model is chosen to be as simple as
possible for the purposes of this investigation.  In contrast with
some of the engineering literature, we restrict the algebraic
complexity in order to retain a physical understanding of the terms in
the equations.  Further refinements are likely to be required in order
to provide an accurate match to a wide range of data.

In comparing a closure model of astrophysical convection with
experimental and numerical results, we face certain difficulties.
Astrophysical convection usually takes place at very high Rayleigh
number, in a highly turbulent regime.  Experiments have been conducted
at very high Rayleigh number but mainly for the Rayleigh--B\'enard
problem in which the flow is dominated by boundary layers, which may
not be relevant in an astrophysical context, or by mean flows not
represented in the closure model.  An alternative system is provided
by the homogeneous Rayleigh--B\'enard problem, which has periodic
boundary conditions in all directions.  This model, however, has
certain peculiarities of its own.  These issues will be addressed in
the sections that follow.

In the remainder of the paper, we develop the closure model first in
the Boussinesq approximation (Section~2) and apply it to the standard
Rayleigh--B\'enard problem (Section~3).  We then consider the
homogeneous Rayleigh--B\'enard system with triply periodic boundary
conditions (Section~4); in this section we also introduce rotation and
discuss the $\Lambda-$effect.  We then adapt the model to the anelastic
approximation for use in stars and other astrophysical flows
(Section~5) and finally draw conclusions (Section~6).  A number of
technical details are covered in the appendices.

\section{Closure model in the Boussinesq system}

\subsection{Basic equations}

In the Boussinesq approximation (e.g. Chandrasekhar 1961) the
equations governing the motion of the fluid are
\begin{equation}
  \partial_iu_i=0,
\end{equation}
\begin{equation}
  \rho_0(\partial_t+u_j\partial_j)u_i=\rho g_i-\partial_ip+
  \rho_0\nu\partial_{jj}u_i,
\end{equation}
\begin{equation}
  \rho=\rho_0\left[1-\alpha(T-T_0)\right],
\end{equation}
\begin{equation}
  (\partial_t+u_i\partial_i)T=\kappa\partial_{ii}T,
\end{equation}
where we have adopted a Cartesian tensor notation.  The dynamical
variables are the velocity $\bu$, the density $\rho$, the pressure $p$
and the temperature $T$.  Quantities regarded as constant in the
Boussinesq approximation are the reference density $\rho_0$, the
reference temperature $T_0$, the coefficient of expansion $\alpha$,
the gravitational acceleration $\bg$, the kinematic viscosity $\nu$,
and the thermal diffusivity $\kappa$.

A simple, static basic state is possible when the temperature is
uniform and the pressure gradient balances gravity, i.e.
\begin{equation}
  T=T_0,
\end{equation}
\begin{equation}
  p=p_0+\rho_0g_ix_i,
\end{equation}
where $p_0$ is a reference pressure.  To examine departures from this
state we define
\begin{equation}
  \Theta=T-T_0,
\end{equation}
\begin{equation}
  \psi={{p-(p_0+\rho_0g_ix_i)}\over{\rho_0}},
\end{equation}
obtaining the governing equations
\begin{equation}
  \partial_iu_i=0,
  \label{boussinesq1}
\end{equation}
\begin{equation}
  (\partial_t+u_j\partial_j)u_i=-\alpha\Theta g_i-\partial_i\psi+
  \nu\partial_{jj}u_i,
  \label{boussinesq2}
\end{equation}
\begin{equation}
  (\partial_t+u_i\partial_i)\Theta=\kappa\partial_{ii}\Theta.
  \label{boussinesq3}
\end{equation}

\subsection{Fluctuations}

We now adopt a standard procedure and separate the dynamical variables
into mean and fluctuating parts, e.g.
\begin{equation}
  u_i=\bar u_i+u_i',\qquad
  \langle u_i'\rangle=0,
\end{equation}
where the angle brackets or the overbar are interchangeably used to  
denote a suitable averaging operation such as
a temporal, spatial or ensemble average.  The mean parts of the
governing equations are
\begin{equation}
  \partial_i\bar u_i=0,
\label{eq:meancont}
\end{equation}
\begin{equation}
  (\partial_t+\bar u_j\partial_j)\bar u_i=-\alpha\bar\Theta g_i-
  \partial_i\bar\psi+\nu\partial_{jj}\bar u_i-\partial_j\bar R_{ij},
\end{equation}
\begin{equation}
  (\partial_t+\bar u_i\partial_i)\bar\Theta=\kappa\partial_{ii}\bar\Theta-
  \partial_i\bar F_i,
\label{eq:meanenergy}
\end{equation}
where
\begin{equation}
  R_{ij}=u_i'u_j'
\end{equation}
is the Reynolds tensor, representing (minus) the turbulent stress, and
\begin{equation}
  F_i=\Theta'u_i'
\end{equation}
represents the turbulent heat flux density.  The problem at hand is to
determine $\bar R_{ij}$ and $\bar F_i$ and thereby close the system
of mean equations. We also introduce the quantity
\begin{equation}
  Q=\Theta^{\prime2},
\end{equation}
representing the temperature variance. 
It should be noted that all three quadratic correlations $R_{ij}$, $F_i$ and
$Q$ will be redefined when we
move on to the (more relevant) anelastic system in which the reference density is non-uniform, but these definitions
are convenient for the Boussinesq system.

The fluctuating parts of the governing equations are
\begin{equation}
  \partial_iu_i'=0,
  \label{divu'}
\end{equation}
\begin{eqnarray}
  \lefteqn{(\partial_t+\bar u_j\partial_j)u_i'+u_j'\partial_j\bar u_i=
  -\alpha\Theta'g_i-\partial_i\psi'+\nu\partial_{jj}u_i'}&\nonumber\\
  &&-\partial_j(R_{ij}-\bar R_{ij}),
  \label{u'}
\end{eqnarray}
\begin{equation}
  (\partial_t+\bar u_i\partial_i)\Theta'+
  u_i'\partial_i\bar\Theta=\kappa\partial_{ii}\Theta'-
  \partial_i(F_i-\bar F_i).
  \label{t'}
\end{equation}
From these we can obtain exact equations for $\bar R_{ij}$, $\bar F_i$
and $\bar Q$ in the form
\begin{eqnarray}
&&(\partial_t+\bar u_k\partial_k)\bar R_{ij}+
  \bar R_{ik}\partial_k\bar u_j+\bar R_{jk}\partial_k\bar u_i\nonumber \\
 && \qquad +
  \alpha(\bar F_ig_j+\bar F_jg_i) - \nu \partial_{kk} \bar R_{ij}
 =-\langle u_i'\partial_j\psi'+u_j'\partial_i\psi'\rangle\nonumber\\
  &&\qquad-
  \langle u_i'\partial_kR_{jk}+u_j'\partial_kR_{ik}\rangle - 2\nu\langle \partial_{k} u_i'\partial_{k }u_j'\rangle,
\end{eqnarray}
\begin{eqnarray}
  \lefteqn{(\partial_t+\bar u_j\partial_j)\bar F_i+
  \bar R_{ij}\partial_j\bar\Theta+\bar F_j\partial_j\bar u_i+
  \alpha\bar Qg_i - \half(\nu + \kappa) \partial_{jj} \bar F_i} &\nonumber\\
  &&\qquad=-\langle\Theta'\partial_i\psi'\rangle-
  \langle\Theta'\partial_jR_{ij}+u_i'\partial_jF_i\nonumber\rangle\\
  &&\qquad + \half(\nu-\kappa) \langle \partial_j (\Theta'\partial_{j}u_i' - u_i' \partial_j \Theta') \rangle\nonumber\\
  &&\qquad - 
  (\nu+ \kappa) \langle \partial_j u_i' \partial_{j}\Theta'\rangle,
\end{eqnarray}
\begin{eqnarray}
 && (\partial_t+\bar u_i\partial_i)\bar Q+2\bar F_i\partial_i\bar\Theta - \kappa \partial_{ii} \bar Q \nonumber \\
&&\qquad =
  -2\langle\Theta'\partial_iF_i\rangle-
  2\kappa\langle (\partial_{i}\Theta')^2 \rangle.
\end{eqnarray}
The left-hand sides of these equations represent the linear
interaction of $\bar R_{ij}$, $\bar F_i$ and $\bar Q$ with the mean
velocity gradient, the mean temperature gradient and the gravitational
field, as well as their diffusion by the microscopic transport coefficients.  There is no difficulty in treating such terms exactly as they
appear.  The right-hand sides of these equations contain difficult
terms of three sorts: those involving correlations with the pressure
fluctuation $\psi'$, those involving triple correlations of
fluctuating quantities, and dissipative terms involving the microscopic
diffusivities $\nu$ and $\kappa$.  These effects can all be regarded
as `non-linear'; although viscous diffusion, for example, is a linear
process, when the Reynolds number is large the viscous terms can be
significant only when a turbulent cascade has forced structure to
appear on the dissipative scales.  None of the terms on the right-hand
sides of these equations can be written in terms of $\bar R_{ij}$,
$\bar F_i$ and $\bar Q$ without further knowledge of the statistical
properties of the fluctuating quantities, such as the spectrum of the
turbulence, which are determined by the non-linear physics of the
turbulent cascade.

\subsection{Proposed closure model}

We therefore attempt to model the system by retaining the exact forms
of the left-hand sides and proposing simple closures for the
right-hand sides, i.e.
\begin{eqnarray}
&& (\partial_t+\bar u_k\partial_k)\bar R_{ij}+
  \bar R_{ik}\partial_k\bar u_j+\bar R_{jk}\partial_k\bar u_i\nonumber\\
&&\qquad+\alpha(\bar F_ig_j+\bar F_jg_i)-\nu \partial_{kk} \bar R_{ij}\nonumber\\
&&\qquad =\sF_{ij}(\bar R_{ij},\bar F_i,\bar Q,\dots),
\label{dtrij}
\end{eqnarray}
\begin{eqnarray}
  \lefteqn{(\partial_t+\bar u_j\partial_j)\bar F_i+
  \bar R_{ij}\partial_j\bar\Theta+
  \bar F_j\partial_j\bar u_i+\alpha\bar Qg_i - \half(\nu + \kappa)\partial_{jj} \bar F_i }&\nonumber\\
  &&=\sF_i(\bar R_{ij},\bar F_i,\bar Q,\dots),
\end{eqnarray}
\begin{equation}
  (\partial_t+\bar u_i\partial_i)\bar Q+2\bar F_i\partial_i\bar \Theta - \kappa \partial_{ii} \bar Q =
  \sF(\bar R_{ij},\bar F_i,\bar Q,\dots),
\label{dtq}
\end{equation}
where the quantities $\sF$ are non-linear tensorial functions of their
arguments.  The dots represent the parameters of the problem, on which
the functions $\sF$ may depend.

A simple example of such a model is
\begin{eqnarray}
&& (\partial_t+\bar u_k\partial_k)\bar R_{ij}+
  \bar R_{ik}\partial_k\bar u_j+\bar R_{jk}\partial_k\bar u_i\nonumber\\
&&\qquad+
  \alpha(\bar F_ig_j+\bar F_jg_i)
 - \nu \partial_{kk} \bar R_{ij}\nonumber\\
&&\qquad=- \frac{C_1}{L}\bar R^{1/2}\bar R_{ij} -
  \frac{C_2}{L}\bar R^{1/2}
  (\bar R_{ij}-{\textstyle{{1}\over{3}}}\bar R\delta_{ij}) - \nu \frac{C_{\nu}}{ L^2} \bar R_{ij},\nonumber\\
\label{eq:Rprop}
\end{eqnarray}
\begin{eqnarray}
  \lefteqn{(\partial_t+\bar u_j\partial_j)\bar F_i+
  \bar R_{ij}\partial_j\bar \Theta+\bar F_j\partial_j\bar u_i+
  \alpha\bar Qg_i - \half(\nu+\kappa) \partial_{jj} \bar F_{i}}&\nonumber\\
  &&=-\frac{C_6}{L}\bar R^{1/2}\bar F_i - \half(\nu + \kappa) \frac{C_{\nu\kappa}}{ L^2} \bar F_{i} ,
\label{eq:Fprop}
\end{eqnarray}
\begin{equation}
  (\partial_t+\bar u_i\partial_i)\bar Q+2\bar F_i\partial_i\bar \Theta - \kappa \partial_{ii} \bar Q =
  -\frac{C_7}{L}\bar R^{1/2}\bar Q - \kappa \frac{C_{\kappa}}{ L^{2}} \bar Q,
\label{eq:Qprop}
\end{equation}
where $R=R_{ii}$ is the trace of the Reynolds tensor, which is twice
the turbulent kinetic energy per unit mass, and $C_1$, $C_2$, $C_6$
and $C_7$ are positive dimensionless coefficients of order unity, of a
universal nature.  (Coefficients $C_3$, $C_4$ and $C_5$ are reserved
for a magnetohydrodynamic extension of the model, see Ogilvie 2003)

The justification for introducing non-linear terms of the above form
is similar to that used in the model of magnetorotational turbulent
stresses originally introduced by Ogilvie (2003). 
The term involving $C_1$ causes a dissipation of
turbulent kinetic energy, and allows for the free decay of
hydrodynamic turbulence.  The term involving $C_2$ redistributes energy among
the components of $\bar R_{ij}$, and corresponds to the tendency of
hydrodynamic turbulence to return to isotropy through the effect of
the pressure--strain correlation. Both are constructed assuming that these
effects occur on a timescale related to the eddy turnover time, $ L / \bar R^{1/2}$, where $L$ is defined
as the typical scale of the largest turbulent eddies.  Terms
$C_6$ and $C_7$, related to the transport of heat, 
are advanced by simple analogy.  The coefficients must satisfy certain conditions 
to ensure the realizability of the model, as discussed in Appendix~A.

The terms proportional to the microscopic diffusion coefficients are introduced 
to allow a modelling of the correlation terms $2\nu\langle \partial_{k} u_i'\partial_{k }u_j'\rangle$, $(\nu+ \kappa) \langle \partial_j u_i' \partial_{j}\Theta'\rangle$ and $2\kappa\langle (\partial_{i}\Theta')^2 \rangle$ at moderate Reynolds number, i.e.
close to the onset of convection.  In such a situation a turbulent cascade does not form and the dissipative terms are proportional to, rather than independent of, the diffusion coefficients. In a similar way, for turbulent shear flows, GO05
proposed to model the momentum diffusion term as
\begin{equation}
2\nu\langle \partial_{k} u_i'\partial_{k }u_j'\rangle \rightarrow \nu \frac{C_\nu}{L^2} \bar R_{ij}
\label{eq:cnumodel}
\end{equation}
on dimensional grounds. Indeed, it is expected that near the onset of convection, most fluid motions will be on the largest scales of the system ($L$). By analogy, we model the other two terms here as 
\begin{eqnarray}
&& (\nu+ \kappa) \langle \partial_j u_i' \partial_{j}\Theta'\rangle \rightarrow \half(\nu+ \kappa) \frac{C_{\nu\kappa}}{L^2} \bar F_i \mbox{  , } \\
&& 2\kappa\langle (\partial_{i}\Theta')^2 \rangle \rightarrow \kappa \frac{C_\kappa}{L^2} \bar Q \mbox{  . }
\label{eq:ckappamodel}
\end{eqnarray}
Therefore the dissipative term in each of equations (\ref{dtrij})--(\ref{dtq}) is modelled by a sum of two terms, one that is independent of the diffusivity and dominates at high Reynolds numbers, and another that is proportional to the diffusivity and dominates at moderate Reynolds numbers.
This completes the justification for the form of the closure model proposed in 
equations (\ref{eq:Rprop}), (\ref{eq:Fprop}) and (\ref{eq:Qprop}).

\section{Rayleigh--B\'enard convection}

\subsection{Model setup}

We now apply the closure model to the problem of Rayleigh--B\'enard convection. 
We consider a horizontally  infinite, plane-parallel system, where 
the bottom plate is located at height $z=0$ and the top plate at 
height $z=h$. The relative temperature of the bottom plate 
is $\bar \Theta = \Delta T$ while that of the top
plate is $\bar \Theta= 0$.

In this setup, we look for statistically steady 
and horizontally homogeneous solutions assuming 
that mean quantities and correlations between fluctuating
quantities vary only with $z$. We also assume that there are no mean 
flows in the system. Equations (\ref{eq:meancont})-(\ref{eq:meanenergy}) 
and (\ref{eq:Rprop})-(\ref{eq:Qprop}) 
reduce to a set of ordinary differential equations (ODEs) 
which can be solved to obtain the temperature profile
$\bar \Theta(z)$ between the two plates, the profiles of the turbulent kinetic 
energy, $\bar R(z) / 2$, and the temperature variance, $\bar Q(z)$ (for example).

By analogy with Prandtl's mixing-length formulation (Prandtl, 1932) 
we set $L$, the size of the largest eddies, to be equal
to the distance to the nearest wall, i.e.\ $L(z)=\min(z,h-z)$ (see GO05 for
applications of the same principle to pipe flows and to Couette--Taylor flows).

It can be shown with little effort that $\bar R_{xy} = \bar R_{xz} 
= \bar R_{yz} = 0$, as well as $\bar F_x = \bar F_y = 0$. The remaining 
set of five second-order ODEs fully characterizes the system: 
\begin{eqnarray}
\label{eq:RBmodeleq}
\nu \frac{\dd^2 \bar{R}}{\dd z^2} 
&=& \nu \frac{C_\nu}{L^{2}} \bar{R}
                             + \frac{C_1}{ L} \bar{R}^{3/2}
                             - 2 \alpha \bar{F}_z g, \nonumber \\
\nu \frac{\dd^2 \bar{R}_{zz}}{\dd z^2} &=& \nu \frac{C_\nu}{L^2} \bar{R}_{zz}
+ \frac{C_1+C_2}{ L} \bar{R}^{1/2} \bar{R}_{zz}
                                   - \frac{ C_2}{3 L} \bar{R}^{3/2}\nonumber \\
                               &&                                       - 2 \alpha \bar{F}_z g, \nonumber \\
\half(\nu + \kappa) \frac{\dd^2 \bar{F}_{z}}{\dd z^2} &=& \half(\nu + \kappa) \frac{C_{\nu\kappa}}{ L^2} \bar{F}_{z}
                                  + \frac{C_6}{ L} \bar R^{1/2}\bar{F}_z \nonumber \nonumber \\
                                &&  - \alpha \bar{Q} g
                                  + \bar{R}_{zz} \frac{\dd \Theta}{\dd z}, \nonumber \\
\kappa \frac{\dd^2  \bar{Q}}{\dd z^2} &=& \kappa \frac{C_\kappa}{ L^2} \bar{Q}
                                  + \frac{C_7}{ L} \bar R^{1/2} \bar{Q} + 
                                  2 \bar{F}_z \frac{\dd \bar{\Theta}}{\dd z}, \nonumber \\
\kappa \frac{\dd^2 \bar{\Theta}}{\dd z^2} &=& \frac{\dd \bar{F}_z}{\dd z},
\label{eq:RBODE}
\end{eqnarray}
where $g = -g_z$. In the case of no-slip boundaries with fixed temperature on each plate as listed above, $\bar R$, $\bar R_{zz}$, $\bar F_z$ and $\bar Q$ are zero on both boundaries. 

\begin{figure}
\epsfig{file=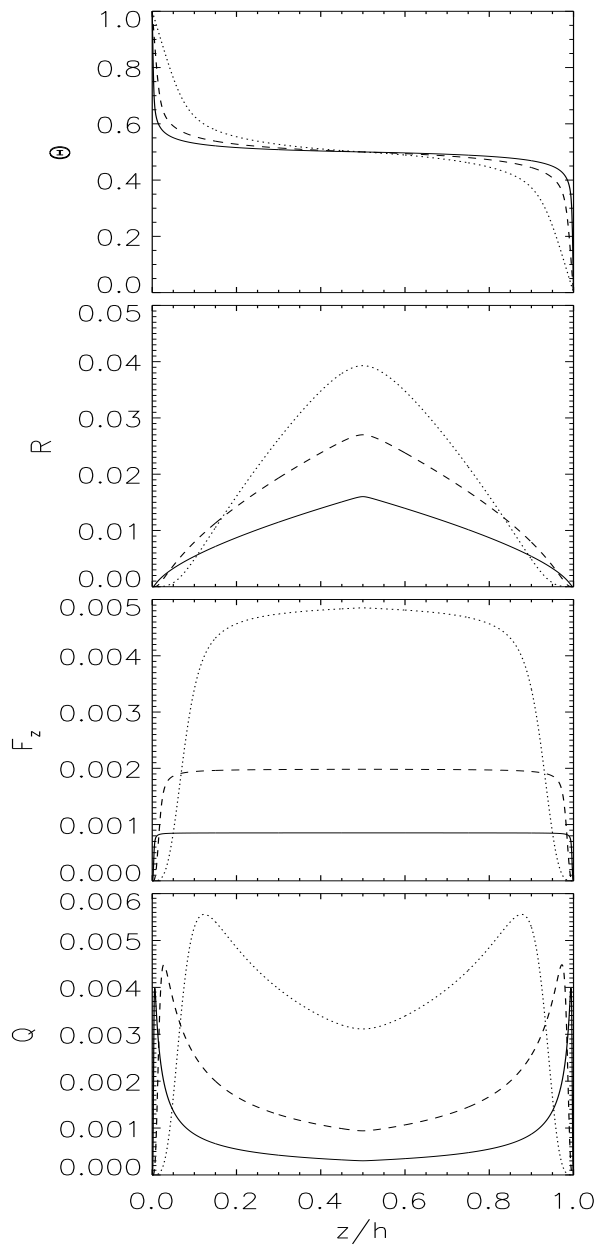,width=8cm,height=17cm}
\caption{Vertical profiles $\bar \Theta(z)$ (in units of $\Delta T$), $\bar R(z)$ 
(in units of $\kappa^2/h^2$), $\bar F_z(z)$ (in units of $\kappa \Delta T/h$) and 
$\bar Q(z)$ (in units of $(\Delta T)^2$) for $\mathrm{Ra}=10^6$ (dotted line), 
$\mathrm{Ra}=10^8$ (dashed line) and  $\mathrm{Ra}=10^{10}$ (solid line). 
In all cases, $\mathrm{Pr} = 1$. }
\label{fig:raplots}
\end{figure}

This system of ODEs with associated boundary conditions can be solved with a two-point boundary-value solver. Typical solutions are shown in Fig.~\ref{fig:raplots} for various Rayleigh numbers, defined here as
\begin{equation}
{\rm Ra} =  \frac{\alpha g h^3 \Delta T}{\nu \kappa}.
\end{equation}
We set the Prandtl number
\begin{equation}
  \mathrm{Pr}=\frac{\nu}{\kappa}
\end{equation}
to~$1$ for the purposes of illustration.
Note the appearance of the characteristically flat temperature profile between 
the two plates as Ra $\rightarrow \infty$ and of the thin 
thermal boundary layers. We now study in more detail the structure of the 
solution. 

\subsection{Universal profile of convection from a wall}
\label{subsec:blbehavior}

As in the case of shear flows past a wall (see GO05), we can derive
a universal profile for convection away from a wall. Let us consider a 
semi-infinite domain $z>0$, in which case $L = z$, and
let $F_0$ be the convective heat flux through the system. 
We define dimensionless variables via
\begin{eqnarray}
&& z = \left[ \frac{\kappa^2 \nu}{\alpha g F_0} \right]^{1/4} \eta\mbox{    ,   }  \nonumber \\
&& \bar F_z = F_0 f(\eta), \qquad  \bar Q = \left[ \frac{F_0^3 \nu}{\alpha g \kappa^2} \right]^{1/2} q(\eta),\nonumber\\
&&  \bar \Theta - \Delta T =  \left[ \frac{F_0^3 \nu}{\alpha g \kappa^2} \right]^{1/4} \theta(\eta) ,\nonumber \\
&& \bar R_{ij} = \left[ \frac{\alpha g F_0 \kappa^2}{\nu} \right]^{1/2} r_{ij}(\eta) ,
\label{eq:nondim}
\end{eqnarray}
so that the system of equations (\ref{eq:RBODE}) becomes
\begin{eqnarray}
r'' &=& \frac{C_{\nu}}{\eta^2} r + \frac{1}{{\rm Pr}} \frac{C_1}{\eta} r^{3/2} - 2f, \nonumber \\
r''_{zz} &=& \frac{C_{\nu}}{\eta^2} r_{zz} + \frac{1}{{\rm Pr}} \frac{C_1+C_2}{\eta} r^{1/2}r_{zz} -\frac{1}{{\rm Pr}} \frac{C_2}{3\eta} r^{3/2}- 2f, \nonumber \\
\frac{{\rm Pr} + 1}{2} f'' &=& \frac{{\rm Pr} + 1}{2}  \frac{C_{\nu\kappa}}{\eta^2} f + \frac{C_6}{\eta} r^{1/2} f - {\rm Pr}\, q + r_{zz} \theta', \nonumber \\
q'' &=&  \frac{C_{\kappa}}{\eta^2} q + \frac{C_7}{\eta} r^{1/2} q + 2 f \theta' \nonumber, \\
\theta' &=& f-1. 
\end{eqnarray}
The boundary conditions at $\eta = 0$ 
are $r=r_{zz}=f=q=\theta = 0$.

Solutions very close to the wall $(\eta \ll 1)$ satisfy:
\begin{eqnarray}
&& r \mbox{  and  } r_{zz} \propto \eta^{\alpha_{\nu}} \mbox{  with  } \alpha_\nu(\alpha_\nu - 1) = C_\nu \nonumber, \\
&& f \propto \eta^{\alpha_{\nu\kappa}} \mbox{  with  } \alpha_{\nu\kappa}(\alpha_{\nu\kappa} - 1) = C_{\nu\kappa}, \nonumber\\
&& q \propto \eta^{\alpha_{\kappa}} \mbox{  with  } \alpha_{\kappa}(\alpha_{\kappa} - 1) = C_{\kappa}.
\label{sublayerprofile}
\end{eqnarray}
These simple relationships provide an ideal way of calibrating each of 
the three constants $C_\nu$, $C_{\nu\kappa}$ and $C_\kappa$ individually 
(see Section~\ref{s:calibration}), by analysing the power-law behaviour of the 
near-wall profiles of experimental or numerical data. 

Solutions far away from the boundary layer can be expanded as
\begin{eqnarray}
&& r = r_0 \eta^{2/3} + O(\eta^{-2/3}), \nonumber \\
&& r_{zz} = r_{zz0} \eta^{2/3} + O(\eta^{-2/3}), \nonumber \\
&& f = 1 - f_1 \eta^{-4/3} + O(\eta^{-8/3}), \nonumber \\
&& q = q_0 \eta^{-2/3} + O(\eta^{-4/3}),  \nonumber \\
&& \theta = \theta_0 + 3f_1 \eta^{-1/3} + O(\eta^{-5/3}),
\label{eq:powerlaweta}
\end{eqnarray}
where 
\begin{eqnarray}
r_0 &=& \left( \frac{2{\rm Pr}}{C_1} \right)^{2/3},\qquad 
r_{zz0} = \frac{3C_1+C_2}{3(C_1 + C_2)} r_0, \nonumber \\
f_1 &=& \frac{C_6}{\frac{C_1}{C_7} + \frac{3C_1 + C_2}{3(C_1+C_2)} } r_0^{-1/2},\qquad
q_0 = \frac{2 f_1}{C_7 r_0^{1/2} }.
\end{eqnarray} 
However, unlike $r_0$, $f_1$ and $q_0$ 
the constant $\theta_0$ cannot be determined without a numerical calculation of the
boundary-layer solution for $\eta = O(1)$.

The scaling laws obtained for $r$, $f$, $\theta$ and $q$ far from the wall
are expected on dimensional grounds, and recover the well-known solution 
of Priestley (1954). They are analogous to the universal ``log-law'' 
solutions for turbulent shear flows past a wall (e.g. Schlichting, 1979).
By comparing profiles of $r$, $f$ and $q$ with laboratory or numerical
experiments, one can constrain some of the unknown 
coefficients $\{C_i\}$ (see Section~\ref{s:calibration}). 

\subsection{Nusselt--Rayleigh number relationship}
\label{subsec:RaNu}

The heat flux through the system in Rayleigh--B\'enard convection is commonly 
measured by the dimensionless Nusselt number
\begin{equation}
{\rm Nu} = 1 + \frac{h F_0}{\kappa \Delta T } ,
\end{equation}
which compares the total heat flux with the conductive one in the absence of convection.
The universal convection-from-a-wall solution calculated 
in the previous section can be used to derive 
the relationship between the Nusselt number and the Rayleigh number. 

Indeed, by selecting a Rayleigh number we set the relative 
temperature at the midpoint $z=h/2$ to be
$\bar \Theta=\Delta T /2$ which implies through (\ref{eq:nondim}) that 
\begin{equation}
\frac{\Delta T}{2} - \Delta T = \left[ \frac{F_0^3 \nu}{\alpha g \kappa^2} \right]^{1/4} \theta\left(  \left[ \frac{\alpha g F_0}{\kappa^2 \nu} \right]^{1/4} \frac{h}{2} \right)  \mbox{   ,  }
\end{equation}
yielding an equation for the (unknown) constant heat flux $F_0$. 
In dimensionless terms, we have the implicit equation for $\mathrm{Nu}$:
\begin{equation}
\frac{1}{2}[\mathrm{Ra}(\mathrm{Nu}-1)^{-3}]^{1/4}=\theta\left(\half[\mathrm{Ra}(\mathrm{Nu}-1)]^{1/4}\right),
\end{equation}
which can be solved to find $\mathrm{Nu}(\mathrm{Ra})$.
In the limit of very large Rayleigh number
the mid-point of the system is very far from 
the boundary layer, so $\theta\approx\theta_0$ 
which then recovers the standard scaling law (Malkus 1954)
\begin{equation}
{\rm Nu} = 1 + \left( \frac{ {\rm Ra}}{16 \theta_0^4}  \right)^{1/3}  \mbox{   .  }
\end{equation}
The constant $\theta_0$ depends only on Pr and on the closure 
parameters $\{ C_i\}$, but cannot easily be expressed analytically 
in terms of these parameters. 

\subsection{Comparison with data and estimation of the model parameters}
\label{s:calibration}

The aim of this section is to estimate, in a rough sense,
the parameters $\{C_i\}$ by comparing the model predictions 
with numerical simulations and laboratory experiments. 
This approach was successfully used in GO05 on pipe
flow data and Couette--Taylor data, yielding:
\begin{equation}
C_1 \simeq 0.4 \mbox{   ,   }  C_2 \simeq 0.6 \mbox{   ,   } C_\nu \simeq 12 .
\label{eq:C1C2}
\end{equation}
Under the assumption that the closure parameters are 
universal properties of the turbulent cascade, these estimated values 
should also apply to the case of turbulent convection 
without need for re-calibration. The remaining parameters $C_6$, $C_7$, 
$C_{\nu\kappa}$ and $C_{\kappa}$ may then be independently estimated.
In the following sections, we first discuss this assumption 
in the light of known model limitations. We then select 
appropriate experimental datasets and 
use them to constrain the remaining parameters.

\subsubsection{Discussion of the model limitations}

As discussed by Ogilvie (2003) and GO05
the closure model proposed has two intrinsic limitations: 
it ignores some (but not all) of the effects of 
pressure-strain correlations $<u'_i \partial_j \psi' >$, 
and assumes that the effect of all modelled terms 
(such as the triple-correlations 
in (\ref{eq:Rprop})-(\ref{eq:Qprop})) is local both in 
time and space. As a result, it
may poorly represent strongly sheared systems 
or systems where the turbulent eddies exhibit a strong degree of 
spatial or temporal coherence. 

The neglected effects of the pressure-strain correlations are not thought to be important in 
turbulent convection, except in the presence of strong rotation or of 
an externally driven strong mean shear (where the timescale of rotation and shear is comparable to 
that of the convection). The closure should be
well-suited to model convection in stellar interiors, but maybe less so for 
convectively unstable accretion discs. We defer this particular case to subsequent work. 
However, for similar reasons these effects are also likely to be important in pipe flows or 
Couette--Taylor flow, which were used as a basis for calibrating the constants 
$C_1$ and $C_2$ (see GO05). Consequently, the estimates given in (\ref{eq:C1C2})
could be somewhat biased, in particular $C_2$ which contains information on the rate of
return to isotropy. Comparing the model with turbulent convection experiments (see below) can
therefore help refine the estimates for $C_1$ and $C_2$ using more appropriate data. 

As mentioned above, the closure is also less reliable when applied to systems where the turbulence
exhibits coherence over large scales or long timescales. This might pose some problems
when applied to convection in a finite domain, since large-scale coherent plumes which span
the whole system are commonly observed in most cases ranging from Boussinesq 
to fully compressible systems. Comparisons with experiments can help reveal which aspects 
of convective transport are adequately described by the model, and which are not. 

\subsubsection{Available experimental data}

Our application of the closure model to Rayleigh-B\'enard convection in Sections 
\ref{subsec:blbehavior} and \ref{subsec:RaNu} assumes for simplicity  
that the system is horizontally invariant, while all laboratory and numerical 
experiments have a limited horizontal extent. 
The presence, nature and geometry of the side-walls are known to affect 
various properties of the turbulent convection, in particular 
through the generation of large-scale circulations (often called ``wind''). 
This wind influences the overall heat transport properties 
by changing the nature of 
the boundary layers (Castaing et al. 1989; Cioni et al. 1997;
Grossmann \& Lohse 2000, 2001, 2002, 2004). 
It also induces large-scale 
horizontal inhomogeneities, so that the measured vertical profiles 
of mean quantities and higher-order moments may vary 
with position (Maystrenko, Resagk \& Thess 2007). 
While our formalism can in principle be applied to finite
geometries and self-consistently model the effect of large-scale 
flows, such an extension is beyond the scope
of the present paper. 

In order to minimize the effect of side-walls
we restrict the model comparison to experimental setups 
with very large aspect ratios (defined as the ratio of the horizontal
to vertical extent of the domain, and denoted as $\Gamma$). 
There are a few large aspect ratio, high Rayleigh number
experimental studies which provide measurements of the Nusselt number.
Of particular interest are results of  
F\"unfschilling et al. (2005) for convection 
in water (Pr = 4.38) in a cylindrical
enclosure of aspect ratio up to $\Gamma= 6$, 
for Ra up to a few times $10^{10}$.
Niemela \& Sreenivasan (2006) provide similar information 
for convection in Helium (0.7 $<$ Pr $<$ 8) 
in a cylindrical container with $\Gamma = 4$, 
for Rayleigh numbers between $10^8$ and $10^{13}$.  
Finally, the Ilmenau barrel experiments of 
DuPuits, Resagk \& Thess (2007) provide Nu(Ra) for convection in 
air (Pr = 0.7) in a cylindrical enclosure with variable aspect ratio up 
to 11.3, for Rayleigh numbers up to a few times $10^8$ (in the case of the
largest aspect ratio). 

By contrast, only very few large aspect ratio experimental measurements of 
the boundary-layer profiles of velocity and temperature correlations 
(such as $\bar R_{ij}$, $\bar F_i$ or $\bar Q$) have been reported. 
The largest aspect ratio experiments available ($\Gamma =11.3$) 
with fully resolved boundary layer profiles are presented by 
DuPuits, Resagk \& Thess (2007) although the 
data provided is limited to the mean and rms temperature profiles.

Taking a different approach, direct numerical experiments are a powerful tool 
for ``idealized'' experiments. Horizontally periodic simulations minimize
the effect of side-walls (although retain a finite aspect ratio) and permit 
resolved and precise measurements of all desired mean and fluctuating
quantities within the flow. The main drawback is the limited range 
of parameter space for which resolved simulations can be run 
(typically, Ra $< 10^8$ for large aspect ratio simulations at Pr $= O(1)$). 

\begin{figure}
\epsfig{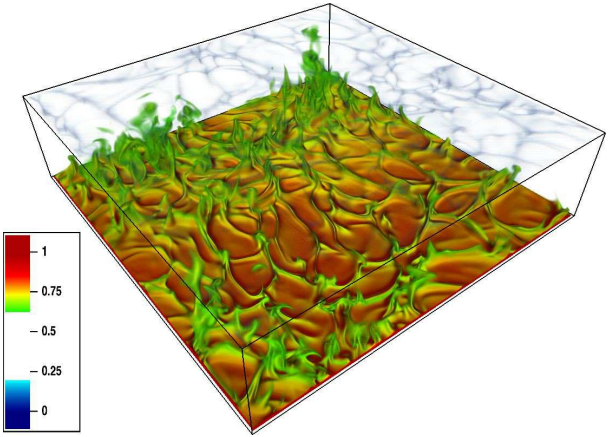}
\caption{Volume-rendered visualization of the temperature field in 
our numerical simulation of Rayleigh-B\'enard convection for 
${\rm Ra} = 2.1 \times 10^7$, and $\mathrm{Pr} = 1$. The system is doubly-periodic
in the horizontal direction, with aspect ratio 4, 
and has no-slip boundary conditions at the top and bottom boundary. The colour and 
opacity scheme has been selected to emphasize structures near the lower boundary layer.}
\label{fig:RBeyecandy}
\end{figure}

For these reasons, we use a combination of
experimental data (DuPuits, Resagk \& Thess 2007) and numerical 
simulations to calibrate the remaining model parameters. 
Our numerical simulations are all run for Pr = 1, 
in a horizontally periodic domain with aspect ratio $L_x/L_z = L_y/L_z = 4$, 
using a spectral method briefly described in Appendix~B. The largest
Rayleigh number achieved in this case is Ra $= 2.1 \times 10^7$. 
Figure \ref{fig:RBeyecandy} shows a typical snapshot of the results, in this
parameter regime, for the temperature field for example. The 
results of the simulations are globally consistent with those
of Hartlep (PhD thesis, 2005, G\"ottingen). 

\subsubsection{Near-wall profiles and estimation of $C_\nu$, $C_\kappa$ and $C_{\nu\kappa}$}

Very close to the wall ($\eta \ll 1$), the closure model solutions for 
the normalized
correlations $r$, $r_{zz}$, $f$, $q$ and $\theta$ are well approximated 
by power laws, 
as described in equation~(\ref{sublayerprofile}). These relationships can 
be compared with data and provide a simple 
way of individually estimating each of the model constants 
$C_\nu$, $C_\kappa$ and $C_{\nu\kappa}$ 
from laboratory or numerical experiments.

\begin{figure}
\epsfig{file=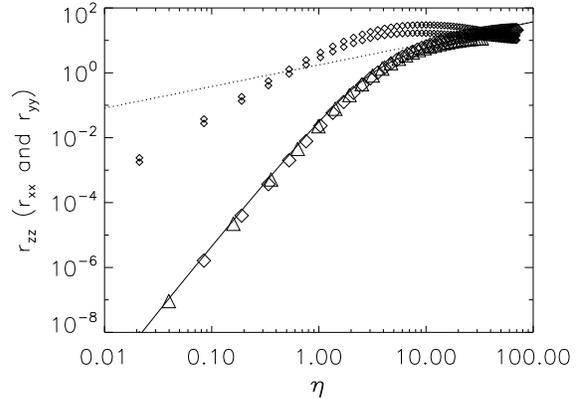,width=8cm}
\caption{Comparison of the universal ``convection from a wall''
solution  with numerical data for the dimensionless Reynolds stress
components  $r_{zz}$, $r_{xx}$ and $r_{yy}$. The large symbols
represent  $r_{zz}(\eta)$  for Ra = $2.1 \times 10^6$ (triangles)  and
$2.1 \times 10^7$ (diamonds). The two sets of smaller diamonds show
$r_{xx}(\eta)$ and $r_{yy}(\eta)$ for the case where Ra= $2.1 \times
10^7$. Note that theoretically these should be lying on the same curve
-- the difference can be attributed to limited statistics.  In all
cases $\mathrm{Pr}=1$. The dotted line shows the asymptotic solution
$r_{zz} = r_{0zz} \eta^{2/3}$ using the value of $C_1$ estimated by
GO05, while the solid line shows a numerical integration of the full
universal profile, for our estimated parameter values as listed in
(\ref{eq:C1C2}),  (\ref{eq:Cd}), and (\ref{eq:C6C7}). }
\label{fig:rzzcal}
\end{figure}

\begin{figure}
\epsfig{file=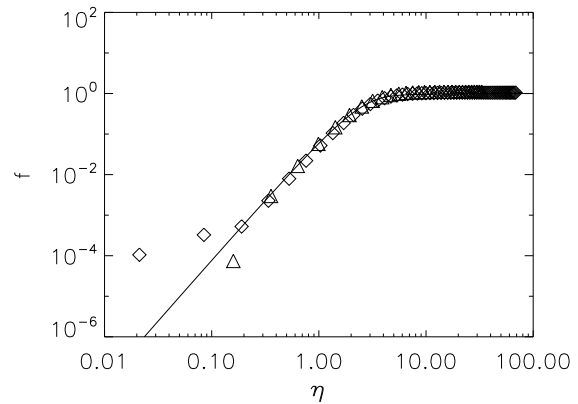,width=8cm}
\caption{Comparison of the predicted dimensionless convective heat
flux profile $f(\eta)$ with our numerical data.
The symbols have the same meaning as in Fig.~\ref{fig:rzzcal}. Note
that the scatter for $\eta < 0.1$ comes from imperfect statistics very
close to the wall.  This plot was used to fit $C_{\nu\kappa}$ to
capture the near-wall solution correctly. The solid line  shows a
numerical integration of the full universal profile, for our estimated
parameter values as listed in  (\ref{eq:C1C2}),  (\ref{eq:Cd}), and
(\ref{eq:C6C7}).}
\label{fig:fzcal}
\end{figure}

Comparisons of (\ref{sublayerprofile}) with the experimental  
near-wall profile for $r_{zz}(\eta)$, $f(\eta)$ and $q(\eta)$
yield slopes $\alpha_\nu$ close to 4 
(see Fig.~\ref{fig:rzzcal}), $\alpha_{\nu\kappa}$ close to 3 (see Fig.~\ref{fig:fzcal}), 
and $\alpha_\kappa$ close to 2 (see Fig.~\ref{fig:qcal}).
Note that while the amplitude of the power-law observed in 
the near-wall profile for $q(\eta)$ is seen to depend on the experiment considered, 
the slope $\alpha_\kappa$ appears to be universal. 
We then adopt the following values for the constants $C_\nu$, $C_{\nu\kappa}$ and $C_\kappa$:
\begin{eqnarray}
&& C_\nu = 12\pm 1 \mbox{  ,  }  \nonumber \\
&& C_{\nu\kappa} = 6 \pm 0.5 \mbox{  ,   }  \nonumber \\
&& C_{\kappa} = 2 \pm 0.2 \mbox{  .   } 
\label{eq:Cd}
\end{eqnarray}
Given the experimental and model uncertainties, these values
and their errorbars should be thought of as rough estimates rather
than precise calibrations.

It is comforting to note that this independent comparison recovers  the
value of $C_\nu$ found by GO05. Moreover, we find that 
within  fitting errors $C_{\nu\kappa}
\simeq (C_\nu C_\kappa)^{1/2}$.  Given the quantities 
modelled by the associated diffusive terms (see equations
(\ref{eq:cnumodel})--(\ref{eq:ckappamodel})), this result  is not
entirely surprising.

On the other hand, Fig. \ref{fig:rzzcal} reveals an important  caveat
of the closure model when applied to Rayleigh-B\'enard convection.
The universal solution for the two horizontal stress components 
$r_{xx}(\eta)$ and
$r_{yy}(\eta)$ can easily be deduced from  $r_{xx} = r_{yy} = 0.5 (r -
r_{zz})$.  These horizontal stresses should therefore be identical to
one another and have the same power-law dependence on $\eta$ as $r$
and $r_{zz}$, close to the wall and far from the wall. However,
Fig. \ref{fig:rzzcal} clearly shows that the numerical data is at odds
with the model. We attribute the discrepancy to the presence of
large-scale coherent convective plumes in the system, which span the
entire domain and create strong horizontally correlated fluctuations
as they crash against each boundaries. As a result, the fluid in the
viscous sublayer is much more strongly anisotropic than predicted.

\subsubsection{Far-field solution and estimation of $C_6$ and $C_7$.}

Fig.~\ref{fig:rzzcal} compares the predicted profile for
$r_{zz}(\eta)$ with data from our numerical simulations. The dotted
line shows the model prediction for the solution far from the wall
$r_{zz} = r_{zz0} \eta^{2/3}$. Note that $r_{zz0}$ depends only on two
numbers, the Prandtl number (which is known)  and the model parameter
$C_1$. It is reassuring to see that the value of $C_1$ estimated by
GO05 from wall-bounded shear flow data adequately
fits the far-from wall solution for $r_{zz}$ in this convection
problem.

\begin{figure}
\epsfig{file=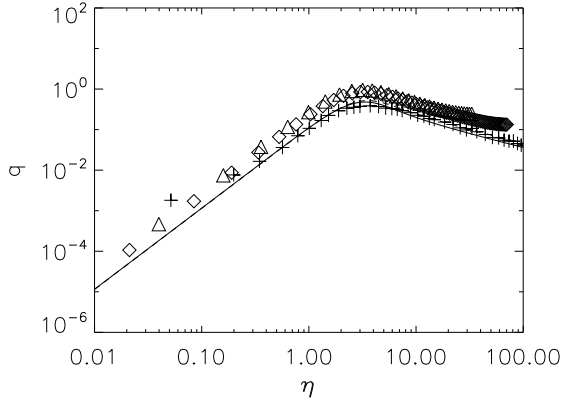,width=8cm}
\caption{Comparison of the predicted dimensionless temperature
variance $q$ with experimental and numerical data. The open  symbols
represent the results of our numerical simulations (Pr = 1) for Ra =
$2.1 \times 10^6$ (triangles) and $2.1 \times 10^7$ (diamonds).  The
plus symbols are  experimental data from DuPuits, Resagk \& Thess
(2007) for Ra =  8.14$\times 10^8$ for air (Pr = 0.7) in a cylindrical
box at aspect  ratio 11.3. The discrepancy between the numerical
solutions  and the experimental data is attributed to the difference
between periodic side-walls and impermeable side-walls.
The near-wall solution was used to fit $C_\kappa$ while the
far-from-the-wall data was used to provide a constraint between $C_6$
and $C_7$. The solid line show a numerical integration of the full
universal  profile as in Figs. \ref{fig:rzzcal} and \ref{fig:qcal} for
Pr = 1. }
\label{fig:qcal}
\end{figure}



\begin{figure}
\epsfig{file=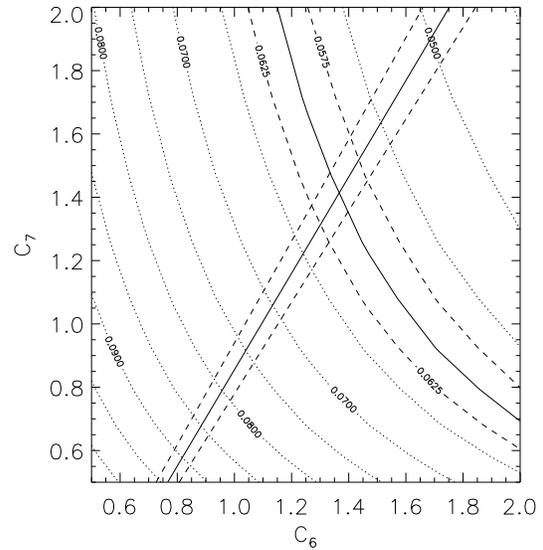,width=8cm}
\caption{Calibration of the constants $C_6$ and $C_7$. The straight
lines  show the relationship between $C_6$ and $C_7$ when the constant
$q_0$  is equal to 0.95 (solid line), 0.9 or 1.0 (dashed lines, top
and bottom respectively).  The curves show the value of $K$ in the
relationship  ${\rm Nu} \sim 1 + K {\rm Ra}^{1/3}$, as predicted by
numerical integrations of the  closure model equations
(\ref{eq:RBmodeleq}) for no-slip boundary conditions. The area  marked
by the intersection of the 4 dashed lines, and centred on the point
where the  two solid lines cross, provides estimates for $C_6$ and
$C_7$.}
\label{fig:C6C7}
\end{figure}

The universal profiles away from the wall listed in equation
(\ref{eq:powerlaweta}) can also be used in conjunction with numerical and laboratory
experiments to constrain $C_6$ and $C_7$.
These constants are unfortunately difficult to extract
directly from our numerical  simulations. The highest Rayleigh number
available (Ra = 2.1 $\times 10^7$) only has a short asymptotic ($\eta
\gg 1$) range, so that  estimates of $C_6$ and $C_7$ from these
datasets are unreliable\footnote{This statement can be verified using
a simple test problem in which artificial data are created {\it using}
the closure model, and then used blindly to reconstruct $C_6$ and
$C_7$.}. The rms temperature data measured in various laboratory experiments at
higher Rayleigh number provides a more adequate  point of comparison.
We use the rms temperature data of the  highest aspect ratio
experiments of DuPuits, Resagk \& Thess (2007), for Ra = $8.14 \times
10^8$. This dataset exhibits a significant asymptotic range, with a
power law close to the one predicted by the closure model  ($q \sim
q_0 \eta^{-2/3}$). Fitting the data yields $q_0 = 0.95 \pm 0.05$,
which provides a first constraint between $C_6$ and $C_7$ (see
Fig.~\ref{fig:C6C7}).  Note that other datasets (from Maystrenko,
Resagk \& Thess, 2007, for example) are generally consistent with this
estimate for $q_0$.

A second constraint between $C_6$ and $C_7$ is obtained by comparing
the model predictions with experimental measurements of $\mathrm{Nu}(\mathrm{Ra})$.
The closure model implies that ${\rm Nu} = 1 + K {\rm Ra}^{1/3}$ where
the constant $K$ is a function of the model parameters (and the
Prandtl number). The data from F\"unfschilling et al. (2005),
Niemela \& Sreenivisan (2006) and  DuPuits, Resagk \& Thess (2007) are
reasonably well approximated by taking  $K=0.06 \pm 0.003$. Variations
of $K$ with Prandtl number, for the range of  experiments discussed,
are within the errorbars.  Given that $C_1$, $C_2$,  $C_\nu$,
$C_\kappa$  and $C_{\nu\kappa}$ are now known, for fixed Prandtl
number, fitting $K$ provides a  unique relationship between $C_6$ and
$C_7$, as seen in Fig.~\ref{fig:C6C7}.

By combining these two constraints, we conclude that a good fit to the
data can be obtained with
\begin{equation}
C_6  = 1.4 \pm 0.1 \mbox{   ,   } C_7 = 1.4 \pm 0.1.
\label{eq:C6C7}
\end{equation}
The values for $\{C_i\}$ quoted in equations (\ref{eq:C1C2}),
(\ref{eq:Cd}), and (\ref{eq:C6C7}) form from here on our selected set
of parameters. These values are to be taken as  indicative estimates,
rather than precise calibrations. We note that the  parameters derived
do satisfy realizability (see Appendix~A). The solid lines shown in
Figs. \ref{fig:rzzcal}, \ref{fig:fzcal} and \ref{fig:qcal}  are the
universal boundary layer profiles calculated using these parameters,
and are seen to fit all datasets (except for $r_{xx}$ and $r_{yy}$, as
discussed above) satisfactorily.

\begin{figure}
\epsfig{file=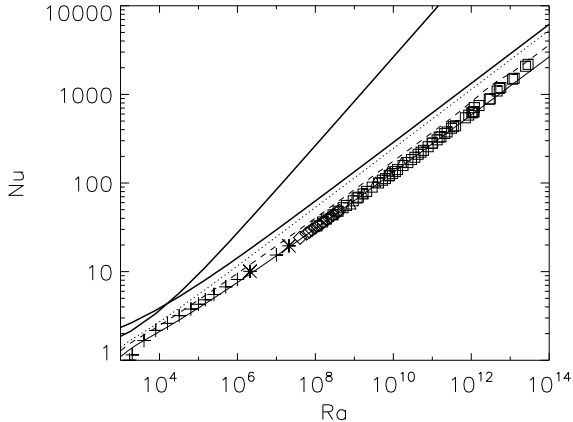,width=8cm}
\caption{Comparison of the model predictions with data for the Nusselt
number as  a function of the Rayleigh number. The square symbols are
experimental data from  Niemela \& Sreenivisan (2006) with Pr $\simeq
1$ (Helium), and aspect ratio  $\Gamma= 4$. The diamond symbols are
the data from F\"unfschilling et al. (2005)  with $\Gamma= 6$,
Pr $= 4.38$ (water). The triangles are data from  DuPuits et al.
(2007), for $4\le \Gamma \le  11.3$, for Pr = 0.7 (air).  The plus
symbols are numerical data from Hartlep et al. (2007), with
$\Gamma = 10$ and for Pr = 0.7. Finally, the star symbols are our own
numerical simulations. The various thin lines shows the closure model
predictions for fiducial values of the parameters $C_i$, for Pr = 1
(solid line), Pr = 4.38 (dashed line) and Pr $\rightarrow \infty$
(dotted line). In addition, the two thick solid lines correspond to
strict upper bound limits: the Nu$=1+0.133\,\mathrm{Ra}^{1/3}$  line
is a strict upper bound obtained by Ierley, Kerswell \&  Plasting
(2006) for Rayleigh--B\'enard convection at infinite  Prandtl number,
while the Nu $= 1+0.0264\,\mathrm{Ra}^{1/2}$  line is a strict upper
bound obtained by Plasting \& Kerswell (2003)  for Rayleigh--B\'enard
convection at arbitrary (finite) Prandtl number. }
\label{fig:RaNu}
\end{figure}

Fig.~\ref{fig:RaNu} compares our closure model prediction for the
Nu(Ra)  relationship, using the estimated parameters, with various
available datasets for large aspect ratio experiments ($\Gamma \ge
4$). It also shows (as dashed lines), for comparison, strict upper
bounds obtained by Plasting \& Kerswell (2003) and by Ierley, Kerswell
\&  Plasting (2006) for transport by convection at finite and infinite
Prandtl numbers respectively. It is reassuring to see that the
$\mathrm{Pr}\rightarrow\infty$ prediction from our own closure model
remains below the strict upper bound for the same limit.

In conclusion, our model successfully reproduces
most measurable features pertaining to laboratory and numerical
experiments of  Rayleigh-B\'enard convection, for reasonable values of
the model  parameters $\{C_i\}$. Furthermore, comparison of the
estimated parameter  values across a range of experiments in
other systems  (pipe flows, Couette--Taylor flows) shows that they are
indeed of a universal nature, a results which can only increase
confidence in our approach.  

\section{Homogeneous Rayleigh--B\'enard convection}
\label{A minimal model system}

\subsection{Introduction}

Another system that is of interest, and possibly more relevant to
astrophysical applications, consists of an unbounded layer in which
there is no mean flow, while the mean temperature gradient
$\nabla\bar\Theta$ is uniform and parallel to the  gravitational
acceleration (taken to be in the $z$-direction).  The
evolution of perturbations to this mean state can be described  by the
following set of Boussinesq equations:
\begin{eqnarray}
\frac{\partial \bu'}{\partial t} + \bu' \cdot \nabla \bu' = - \alpha
\Theta' g_z\,\be_z -\nabla\psi' + \nu \nabla^2 \bu' \mbox{  ,  }
\nonumber \\ \frac{\partial \Theta'}{\partial t} + \bu' \cdot \nabla
\Theta' + u_z' \frac{\dd \bar \Theta}{\dd z} = \kappa \nabla^2 \Theta'
\mbox{  ,  } \nonumber \\ \nabla \cdot \bu'= 0 \mbox{  ,  }
\label{eq:HRBorig}
\end{eqnarray}
where all perturbations are triply periodic, as for example
\begin{eqnarray}
\bu'(x,y,z,t) &=& \bu'(x + L_x,y,z,t) \nonumber\\ &=&
 \bu'(x,y+L_y,z,t) \nonumber\\ &=&  \bu'(x,y,z+L_z,t).
\end{eqnarray}
This model setup is now commonly referred to as Homogeneous
Rayleigh--B\'enard (HRB) convection (Borue \& Orszag 1997; Lohse \&
Toschi 2003; Calzavarini et al.\ 2005; Calzavarini et al.\ 2006).
While this system cannot be studied using laboratory experiments, it
lends itself relatively easily to numerical experimentation using
spectral methods in particular. The relevant dimensionless parameters
are the Prandtl number ${\rm Pr}=\nu/\kappa$, the  Rayleigh number, now defined as
\begin{equation}
{\rm Ra}=\frac{\alpha g_z L_z^4\frac{\dd \bar\Theta}{\dd
z}}{\nu\kappa},
\end{equation}
and the aspect ratio(s) $\Gamma =L_{x,y}/L_z$.

The microscopic diffusivities are included in the original equations
(\ref{eq:HRBorig}) to regularize the system by allowing for
dissipation and irreversibility. However, note that the periodic
boundary  conditions forbid the formation of boundary layers,  so it
may be conjectured that  the macroscopic statistical properties of the
turbulent convection should be well defined and independent  of $\nu$
and $\kappa$ in the limits ${\rm Ra}\to\infty$ (Spiegel 1971).
Furthermore, we may expect the  turbulence to be statistically steady
and homogeneous, although anisotropic.  These properties have been
argued to be more  relevant to convection in astrophysical systems
than standard  Rayleigh--B\'enard convection. The HRB model may
therefore provide  a suitable local model of convection deep inside a
star or planet.

On dimensional grounds, the rms turbulent velocity, for example, must
be expressible in the form
\begin{equation}
  \langle u^{\prime2}\rangle^{1/2}= \left(\alpha g_z{{{\rm
  d}\bar\Theta}\over{{\rm d}z}}\right)^{1/2}L_z\, f({\rm Ra},{\rm
  Pr},\Gamma),
\end{equation}
where $f$ is a dimensionless function.  According to the discussion
above,  $f$ should tend to a non-zero function of $\Gamma$ alone in
the limit $\mathrm{Ra}\to\infty$.  It is tempting to conjecture that
$f$ also becomes independent of $\Gamma$ in the limit of large aspect
ratio, $\Gamma\to\infty$.  This would imply that the vertical
length-scale $L_z$ plays a fundamental role in determining the
saturation level of the turbulent convection, presumably by limiting
the size of coherent structures (`eddies').   For convection deep
inside a star or planet, it is the pressure scale-height that imposes
a characteristic vertical scale on the turbulence (see
Section~\ref{sec:anelastic});  in the local model, the vertical extent
of the box plays an equivalent role.  In practice, owing to some peculiarities
of the HRB system discussed below, the role of the aspect ratio in the 
behaviour of the solutions is not so straightforward.

\subsection{Closure model for HRB}

\subsubsection{Governing equations}

Applying our closure model to HRB, and noting that all statistical
averages are now independent of position, we obtain the system of ODEs
for the temporal evolution of the second-order  correlations $\bar
R_{ij}$, $\bar F_i$ and $\bar Q$:
\begin{eqnarray}
&&\partial_t\bar R_{xx} = -\frac{C_1+C_2}{L}\bar R^{1/2}\bar R_{xx}+
\frac{C_2}{3L}\bar R^{3/2} ,\nonumber\\ &&\partial_t\bar R_{xy} =
-\frac{C_1+C_2}{L}\bar R^{1/2}\bar R_{xy},\nonumber\\ &&\partial_t\bar
R_{xz}+\alpha\bar F_xg_z =  -\frac{C_1+C_2}{L}\bar R^{1/2}\bar R_{xz}
,\nonumber\\ &&\partial_t\bar R_{yy} =-\frac{C_1+C_2}{L}\bar
R^{1/2}\bar R_{yy}+ \frac{C_2}{3L}\bar R^{3/2}  ,\nonumber\\
&&\partial_t\bar R_{yz}+\alpha\bar F_yg_z =  -\frac{C_1+C_2}{L}\bar
R^{1/2}\bar R_{yz}  ,\nonumber\\ &&\partial_t\bar R_{zz}+2\alpha\bar
F_zg_z =  -\frac{C_1+C_2}{L}\bar R^{1/2}\bar R_{zz}+
\frac{C_2}{3L}\bar R^{3/2}   ,\nonumber\\ &&\partial_t\bar F_x+\bar
R_{xz}{{{\rm d}\bar\Theta}\over{{\rm d}z}} =  -\frac{C_6}{L}\bar
R^{1/2}\bar F_x ,\nonumber\\ &&\partial_t\bar F_y+\bar R_{yz}{{{\rm
d}\bar\Theta}\over{{\rm d}z}} =  -\frac{C_6}{L}\bar R^{1/2}\bar F_y
,\nonumber\\ &&\partial_t\bar F_z+\bar R_{zz}{{{\rm d}\bar
\Theta}\over{{\rm d}z}}+ \alpha\bar Q g_z = -\frac{C_6}{L}\bar
R^{1/2}\bar F_z  ,\nonumber\\ &&\partial_t\bar Q+2\bar F_z{{{\rm
d}\bar \Theta}\over{{\rm d}z}} =  -\frac{C_7}{L}\bar R^{1/2}\bar Q.
\label{hrb}
\end{eqnarray}
where we have ignored for simplicity contributions from terms
including $C_\nu$, $C_\kappa$ and $C_{\nu\kappa}$ which  do not
contribute to the high-Rayleigh number dynamics of HRB convection.

Note that the resulting equation for $\bar R$ is
\begin{equation}
\partial_t\bar R + 2\alpha\bar F_zg_z =- \frac{C_1}{L}\bar R^{3/2}
\mbox{  ,}
\end{equation}
so that these equations consist of a main system for $(\bar R,\bar
R_{zz},\bar F_z,\bar Q)$, decoupled systems for $(\bar R_{xz},\bar
F_x)$ and $(\bar R_{yz},\bar F_y)$, and prognostic equations for $\bar
R_{xx}, \bar R_{yy}$ and $\bar R_{xy}$.

\subsubsection{Choice of $L$ and consequences for the coefficients $\{C_i\}$}

While selecting $L$ as the distance to the wall is a natural  choice
for wall-bounded convection or shear flows, a different approach must
be used for triply periodic flows. The largest eddy size in this case
is  limited by the horizontal and vertical scales in the box, so that
$L$ can be assumed to be proportional to $\min(L_x,L_y,L_z)$.

It is important to note that the selection of a different $L$ implies
a potential rescaling of  the $\{C_i\}$ coefficients. For example, had we
selected $L = z/2$ in the  wall-bounded case instead of  $L=z$, then
the estimated $C_1$, $C_2$, $C_6$ and $C_7$ would all  be half the
values quoted in Section~\ref{s:calibration}
since these parameters enter the model in the combinations
$C_1/L$, etc. 
Nevertheless, the ratios of any pairs of constants within 
the group $\{C_1,C_2,C_6,C_7\}$ 
should (presumably) be preserved.
Following these considerations, we elect to keep the estimated values
of the $\{C_i\}$ given in  equations (\ref{eq:C1C2}) and
(\ref{eq:C6C7}), and   calibrate instead  the value of the
proportionality constant  $\delta$ in the expression $L= \delta
\min(L_x,L_y,L_z)$.

\subsubsection{High Rayleigh number HRB convection}


A search for non-trivial fixed points of the dynamical system
(\ref{hrb}) (with $\bar R > 0$) reveals  they are the (positive)
solutions of a quartic equation.  In the limit of large Ra 
it can be shown that there is only one positive fixed point with 
\begin{eqnarray}
  &&\bar R_{xx}=\bar R_{yy}=\left({{C_2}\over{C_1+C_2}}\right)
  {{\bar R}\over{3}},\nonumber\\
  &&\bar R_{zz}=\left({{3C_1+C_2}\over{C_1+C_2}}\right)
  {{\bar R}\over{3}},\nonumber\\
  &&\bar R_{xy}=\bar R_{xz}=\bar R_{yz}=0,\nonumber\\
  &&\bar F_z=-{{C_1\bar R^{3/2}}\over{2L(-N^2)}}
  {{{\rm d}\bar\Theta}\over{{\rm d}z}},
  \nonumber\\
  &&\bar F_x=\bar F_y=0,\nonumber\\
  &&\bar Q={{C_1\bar R}\over{C_7(-N^2)}}
  \left({{{\rm d}\bar \Theta}\over{{\rm d}z}}\right)^2,
\label{eq:dimsols1}
\end{eqnarray}
with
\begin{equation}
  \bar R={{2}\over{C_1C_6}}\left[{{C_1}\over{C_7}}+
  {{3C_1+C_2}\over{3(C_1+C_2)}}\right]L^2(-N^2).
\label{eq:dimsols2}
\end{equation}
Note that, in this case, 
\begin{equation}
\bar Q = \frac{2}{C_6 C_7} \left[{{C_1}\over{C_7}}+
  {{3C_1+C_2}\over{3(C_1+C_2)}} \right] L^2 \left({{{\rm d}\bar \Theta}\over{{\rm d}z}}\right)^2 \propto |\nabla \bar T|^2 .
\end{equation}
This solution represents a state of fully developed turbulent convection, which is statistically steady and homogeneous.  The solution exists in the statistically
axisymmetric subspace in which $\bar R_{xx}=\bar R_{yy}$ and $\bar
R_{xy}=\bar R_{xz}=\bar R_{yz}=\bar F_x=\bar F_y=0$, and is stable
with respect to perturbations transverse to this subspace.  It has the
desired properties that the vertical motion is dominant $(\bar
R_{zz}>\bar R_{xx}=\bar R_{yy})$, while the heat flux is purely
vertical and directed down the temperature gradient.  Moreover,
numerical integrations suggest that, where it exists, this state is
stable and universally attracting.

Defining the Nusselt number Nu as the ratio of the total to the conducted
heat flux, 
\begin{equation}
{\rm Nu} = \frac{\bar F_z - \kappa {{{\rm d}\bar\Theta}\over{{\rm d}z}} }{-\kappa {{{\rm d}\bar\Theta}\over{{\rm d}z}} } ,
\end{equation}
we have, in the limit Ra $\gg$ Pr,
\begin{eqnarray}
{\rm Nu} &=& \sqrt{2} C_1 \left[ \frac{1}{C_1 C_6} \left( \frac{C_1}{C_7} + \frac{3C_1 + C_2}{3(C_1+C_2)} \right) \right]^{3/2}\nonumber\\
&&\quad\times ({\rm Pr Ra})^{1/2} \left( \frac{L}{L_z} \right)^2.
\label{eq:RaNuHRB}
\end{eqnarray}
This scaling recovers the ``ultimate turbulence'' regime, 
where the turbulent transport properties are independent of microscopic
diffusivities (Spiegel 1971). Defining the turbulent Reynolds number 
Re as Re = $L \bar R^{1/2}/\nu$, we have
\begin{equation}
{\rm Re} = \left[ {{2}\over{C_1C_6}}\left({{C_1}\over{C_7}}+
  {{3C_1+C_2}\over{3(C_1+C_2)}}\right)\right]^{1/2} \left( \frac{\rm Ra}{\rm Pr} \right)^{1/2} \left( \frac{L}{L_z} \right)^2,
\label{eq:RaReHRB}
\end{equation}
again reproducing the standard scaling for the ultimate regime of convection. 


\subsection{Comparison with numerical experiments}


Numerical simulations of HRB convection were first performed by Borue 
\& Orzag (1997). More recently, Toschi \& Lohse (2003) 
and Calzavarini et al. (2005) performed a range of 
Lattice--Boltzmann simulations in a cubic geometry, 
for various values of the Rayleigh and Prandtl numbers, and report 
on the first evidence for scalings consistent with the ``ultimate regime'' 
of convection, namely ${\rm Nu} \propto ({\rm Ra Pr})^{1/2}$ and 
${\rm Re} \propto \left({\rm Ra }/{\rm Pr} \right)^{1/2} $. 

However, it is now recognized that the dynamics of HRB convection are
more subtle than previously thought. As discussed by Calzavarini 
et al. (2006), simulations at unit aspect ratio 
show huge fluctuations in the 
instantaneous Nusselt and Reynolds numbers arising from the intermittent
or quasi-periodic (depending on Ra) exponential 
growth of so-called ``elevator modes''. These modes are thus named because
they are independent of $z$, and have the peculiar property of being 
exact nonlinear and exponentially growing solutions of the governing 
equations (\ref{eq:HRBorig}). The most unstable mode has a horizontal 
wavelength equal to the larger horizontal dimension
of the box. Hence, the aspect ratio of the system directly influences
the macroscopic solution. 

This phenomenon has a close parallel 
in shearing-box studies of the magnetorotational instability. In that case, 
forcing by a constant velocity gradient plays the role of the 
constant temperature gradient, while perturbations to the background fields 
are also assumed to be triply periodic. This system is unstable to equivalent 
``channel modes'', exact nonlinear and exponentially growing solutions 
of the equations and 
associated periodic boundary conditions (Goodman \& Xu, 1994). In this case, 
it is known that the channel modes are themselves subject to secondary 
shearing 
instabilities which limit their growths. 
However, the existence and growth rates 
of shearing instabilities depend sensitively on aspect ratio: 
they are strongly inhibited in systems where the streamwise
direction is smaller than the cross-stream directions. As a result, 
systems with roughly cubic geometry are dominated by the channel 
modes and are found to have 
very strongly fluctuating large-scale transport properties, but for
larger aspect ratio the fluctuations are much smaller and the channel modes
are inhibited (Bodo et al. 2008). 

\begin{figure}
\epsfig{file=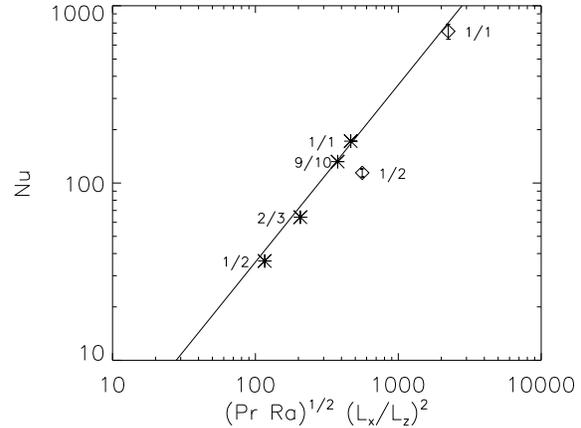,width=8cm}
\caption{Variation of the Nusselt number with rescaled Rayleigh number for 
$\mathrm{Pr}=1$
for homogeneous convection. The diamond symbols show 
the data from our 3D HRB numerical simulations for Ra = $5\times 10^6$ and 
the stars for Ra = $2.16 \times 10^5$. In all cases 
$\mathrm{Pr}=1$. The error bars show the measurement uncertainty due to 
the finite integration time of the simulation. The aspect ratio $\Gamma = L_x/L_z$ of each 
simulation is indicated near the corresponding symbol. The solid line shows
the asymptotic analytical solution (\ref{eq:RaNuHRB}), 
using the values of the parameters $\{ C_i\}$ as listed
in equations (\ref{eq:C1C2}), (\ref{eq:Cd}), and (\ref{eq:C6C7}). A good fit
to the data is found by choosing
$L = L_x/\sqrt{\pi}$.}
\label{fig:HRBRaNu}
\end{figure}

\begin{figure}
\centerline{\epsfig{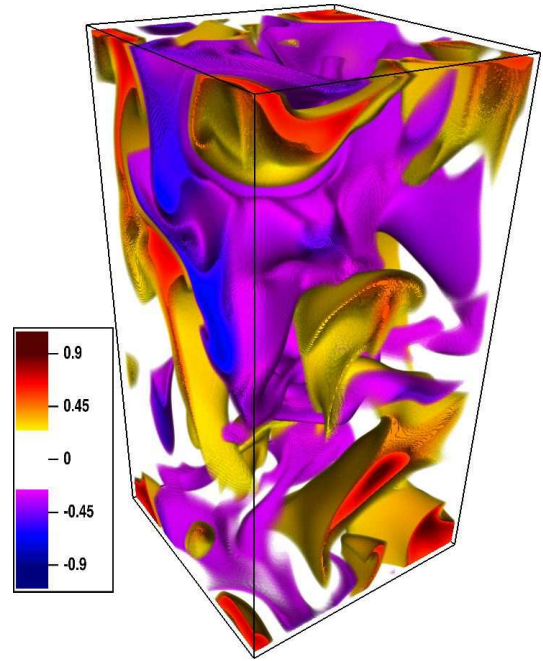}}
\caption{Volume-rendered visualization of the temperature field
for ${\rm Ra} = 5\times 10^6$ and $\mathrm{Pr}=1$
for homogeneous convection in a box of aspect ratio 1/2. Note how, 
even at this high Rayleigh number, the size of the dominant structures
is equal to the box size.}
\label{fig:HRBeyecandy}
\end{figure}

For these reasons, we performed a series of HRB simulations of
various aspect ratios, in order to determine whether the same phenomenon 
occurs, and to provide a better point of comparison for the closure model. 
Appendix~C provides a brief description of the numerical algorithm 
used, and the results are summarized in Fig.~\ref{fig:HRBRaNu}. We studied
5 cases, with $L_x = L_y$ and $L_x/L_z=$1/2, 2/3, 9/10, 1/1 and 4/3. 
In the last case, the elevator modes
continue growing unaffected by perturbations until the code fails, which 
seems to corroborate the premise that the secondary instabilities 
are inhibited 
in wider-than-tall boxes. For $\Gamma < 1$, the measured Nusselt number 
eventually converges to a meaningful average and 
is found to scale as predicted by the closure model, namely proportional 
to (Pr Ra)$^{1/2} \Gamma^2$. 
A good fit with the model predictions is found by selecting 
$L = \delta L_x = L_x/\sqrt{\pi}$. For the purpose of illustration, 
a snapshot of the temperature field for our largest Rayleigh number,
${\rm Ra} = 5 \times 10^6$ (with Pr = 1) and aspect ratio 1/2 is shown
in Figure \ref{fig:HRBeyecandy}.

\subsection{The effect of rotation on homogeneous turbulent convection}
\label{RHRB}

We now consider the effect of rotation on HRB convection, 
where the rotation axis
lies at an angle $\gamma$ from the vertical direction: 
${\bf \Omega} = (0,\Omega \sin\gamma, \Omega \cos\gamma)$. 
In this section it is more convenient to work with dimensionless
variables so we select the following scalings:
\begin{eqnarray}
&& \bar R_{ij} = L^2 \tilde{N}^2 \, \hat R_{ij}, \nonumber \\
&& \bar F_{i} = - \frac{{\rm d} \bar \Theta}{{\rm d} z} L^2 \tilde{N} \, \hat F_{i}, \nonumber \\
&& \bar Q = \left( \frac{{\rm d} \bar \Theta}{{\rm d} z}\right)^2 L^2 \, \hat Q,  \nonumber \\ 
&& \Omega_k = \Omega \, \hat \Omega_k,\qquad g_k = g \, \hat g_k,
\label{eq:nondimrot}
\end{eqnarray}
where for convenience 
$\tilde{N}$ is defined as $\tilde{N}^2 = - N^2$, and is {\it positive} 
when the fluid is convectively unstable. The convective Rossby number 
is then defined as 
\begin{equation}
{\rm Ro}  = \tilde{N}/\Omega.
\end{equation}

Stationary solutions of the closure model far from onset of convection satisfy the following equations: 
\begin{eqnarray}
&& 2\, {\rm Ro}^{-1} (\epsilon_{ikl} \hat R_{lj} + \epsilon_{jkl} \hat R_{li} ) \hat \Omega_k  + \hat g_i \hat F_j + \hat g_j \hat F_i \nonumber \\
&& \qquad = - C_1 \hat R^{1/2} \hat R_{ij} - C_2 \hat R^{1/2} \left(\hat R_{ij} - \frac{\hat R}{3} \delta_{ij} \right),
\end{eqnarray}
\begin{equation} 
- \hat R_{iz} + 2\, {\rm Ro}^{-1} \epsilon_{ijk} \hat\Omega_j \hat F_k + \hat Q \hat g_i = - C_6 \hat R^{1/2} \hat F_i,  
\end{equation}
\begin{equation} 
2 \hat F_{z} = C_7 \hat R^{1/2} \hat Q.
\end{equation}

In the infinite Rossby number limit (equivalently in the non-rotating limit), 
the solution of these equations reduces to the non-dimensional form of (\ref{eq:dimsols1}) 
and (\ref{eq:dimsols2}). 
Should all of the quantities be expanded in terms of the inverse Rossby number as (for example) 
\begin{equation}
\hat{R} = \hat{R}^{(0)} + {\rm Ro}^{-1} \hat{R}^{(1)} + {\rm Ro}^{-2} \hat{R}^{(2)} + \cdots,
\end{equation}
then we find that 
\begin{equation}
\hat{R} = {{2}\over{C_1C_6}}\left[{{C_1}\over{C_7}}+
  {{3C_1+C_2}\over{3(C_1+C_2)}}\right]  + O( {\rm Ro}^{-2}) \mbox{  ,}
\label{eq:norotlimit}
\end{equation}
(and similarly for all diagonal components of $\hat R$). 
Our expressions for the non-diagonal 
terms, to first order, recover the equivalent of the well-known $\Lambda$-effect (see R\"udiger, 1989)
in the coefficient $\hat R_{xz}$:
\begin{eqnarray}
&&\hat R_{xz} = 2 \frac{\hat F_z^{(0)} + C_6 \sqrt{\hat R^(0)} (\hat R_{zz}^{(0)} - \hat R_{xx}^{(0)})}{1-C_6 \hat R^{(0)} (C_1+C_2)} \sin \gamma\, {\rm Ro}^{-1}\nonumber\\
&&\qquad\qquad  + O({\rm Ro}^{-3} ).
\label{eq:leffect}
\end{eqnarray}
The $\Lambda$-effect, as seen in the above equation, describes how rotationally 
constrained turbulent motions can drive differential rotation, through 
the non-diagonal component of the stress-tensor 
$\hat R_{xz}$. As expected from dimensional analysis and geometrical 
arguments, its amplitude scales linearly with $\sin \gamma \,\Omega$.
The other two components $\hat R_{xy} $ and $\hat R_{yz} $ only become 
important for more
rapidly rotating systems as they are both $O( {\rm Ro}^{-3})$. Finally, 
a non-negligible horizontal heat flux is generated in the direction of 
${\bf\Omega} \times \bg$, namely
\begin{eqnarray}
&& \hat F_{x} = 2 \frac{ (\hat R_{zz}^{(0)} - \hat R_{xx}^{(0)}) + (C_1+C_2) \sqrt{\hat R^(0)}  \hat F_z^{(0)} }{1-C_6 \hat R^{(0)} (C_1+C_2)} \sin \gamma\, {\rm Ro}^{-1} \nonumber \\
&&\qquad\qquad + O({\rm Ro}^{-3} )\mbox{   ,} 
\end{eqnarray}
although note that when applied to stellar convection zones, this effect is relevant only for non-axisymmetric
heat transport. The ``latitudinal'' heat flux $\hat F_y$ on the other hand is of higher order in Ro$^{-1}$.

In the opposite limit of very low Rossby number 
(the rapidly rotating limit) an expansion in powers of Ro reveals that 
\begin{equation}
\hat{R} = {{2 \cos^2 \gamma }\over{C_1C_6}}\left[{{C_1}\over{C_7}}+
  {{3C_1+C_2}\over{3(C_1+C_2)}}\right]  + O({\rm Ro})  \mbox{   , }
\label{eq:largerotlimit}
\end{equation}
so that the rms velocity is reduced by a factor $\cos \gamma$ 
compared with the non-rotating case. Note, however, how the expected
reduction (and potential suppression) of the convective heat flux 
in rapidly rotating systems where gravity is {\it aligned} 
with the rotation axis (Chandrasekhar, 1961) so that $\gamma = 0$ 
is not captured by this closure model. This problem, which was identified
by Miller \& Garaud (2007), can presumably be attributed to the 
incomplete modeling of the effects of the
pressure-strain correlations which are known to play 
an important role in the limit of rapid rotation. It is therefore likely 
that these effects also 
cause our model to yield inaccurate predictions for $\gamma \neq 0$ in the same limit.
A full resolution of the issue must eventually involve the 
derivation of a better closure for the pressure-strain correlation terms. 
For completeness note that in this limit 
the model predicts that a significant heat flux is 
carried horizontally along $\be_y$, with amplitude 
$\hat{F}_y = \tan \gamma \, \hat{F}_z$, and that 
\begin{equation}
\hat{R}_{yz} = \frac{C_1}{C_1+C_2} \sin\gamma \cos\gamma \, \hat{R}  + O({\rm Ro})  \mbox{  ,}
\end{equation}
while $\hat{R}_{xy}$ and $\hat{R}_{xz}$ are both $O($Ro$)$.

\begin{figure}
\epsfig{file=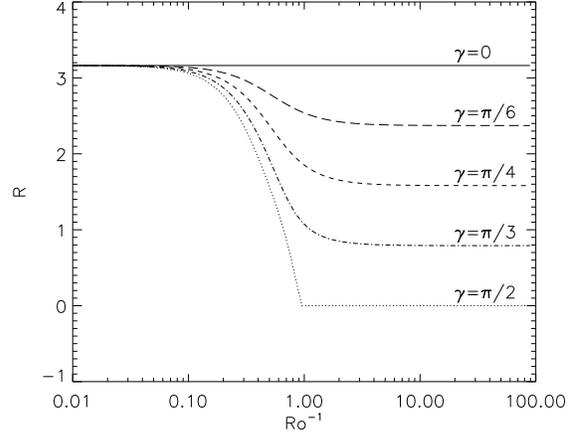,width=8cm}
\caption{Variation of $\hat R$ with Ro$^{-1}$, for various values of $\gamma$, for values of the $\{C_i\}$ parameters given in (\ref{eq:C1C2}) and (\ref{eq:C6C7}). The Ro$^{-1}\rightarrow 0$ and Ro$^{-1}\rightarrow \infty$ asymptotes satisfy equations (\ref{eq:norotlimit}) and (\ref{eq:largerotlimit}) respectively.}
\label{fig:rvsro}
\end{figure}

\begin{figure}
\epsfig{file=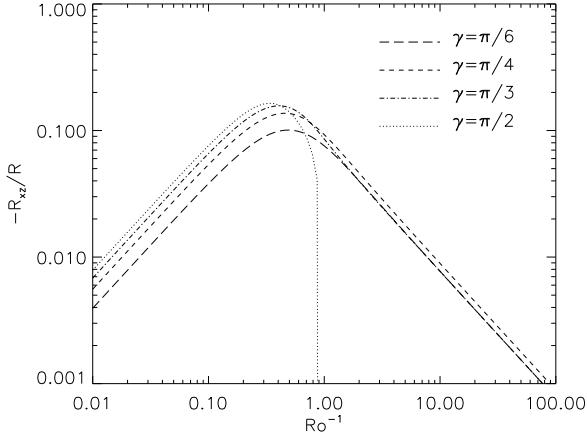,width=8cm}
\caption{Variation of $-\hat R_{xz}/\hat R$ with Ro$^{-1}$, for various values of $\gamma$, for values of the $\{C_i\}$ parameters given in (\ref{eq:C1C2}) and (\ref{eq:C6C7}). Note that $\hat R_{xz}/\hat R \propto \Omega$ for low rotation rates, and to $\Omega^{-1}$ for large rotation rates.}
\label{fig:rxzoverrvsro}
\end{figure}

Fig.~\ref{fig:rvsro} shows the variation of the normalized $\hat R$ 
as a function of both $\gamma$ and Ro$^{-1}$,  while Fig.~\ref{fig:rxzoverrvsro}
shows the variation of the normalized $-\hat R_{xz}/\hat R$ 
as a function of both $\gamma$ and Ro$^{-1}$, illustrating the dependence of the $\Lambda$-effect on 
both parameters as predicted by our model. 

\subsection{Comparison with previous second-order models}

We now compare our findings with the commonly used model for 
convective stresses originally proposed by 
R\"udiger \& Kitchatinov (1993) and later extended by R\"udiger et al. 
(2005, Ral05 hereafter). Note that the related theory
of Kitchatinov \& R\"udiger (2005) relies on the presence 
of a background density stratification to explain the $\Lambda$-effect. As such it 
is not an appropriate point of comparison for our Boussinesq calculation.  

R\"udiger \& Kitchatinov (1993) and Ral05
assume the presence of a ``background'' turbulence caused 
by a given (unspecified) forcing mechanism, which, in the absence of 
rotation, is described by an eddy
turnover time $\tau$, a mixing length $l$ and a turbulent diffusivity $\nu_{\rm t}=l^2/\tau$.
This background turbulence also may also have some degree of 
anisotropy, controlled by the parameter $a$ defined in our notation as
\begin{equation}
a = \frac{\bar R_{xx}^{(0)} +   \bar R_{yy}^{(0)} - 2\bar R_{zz}^{(0)} }{\bar R_{zz}^{(0)} }\mbox{  , }
\end{equation}
where the superscript $(0)$ denotes turbulent 
quantities of the non-rotating system. Note how $a=0$ for isotropic turbulence.  

Ral05 show how the presence of rotation (where the rotation 
axis lies at an angle $\gamma$ from the vertical) modifies the background turbulence, an effect which  
gives rise to non-diagonal components in the stress tensor. They
argue that this phenomenon is controlled 
by the Coriolis number $\Omega^*$ defined as
\begin{equation}
\Omega^* = 2\tau \Omega \mbox{  . }
\end{equation}
Their eddy turnover time $\tau$ is naturally 
related to $L/\sqrt{R^{(0)}}$ in our notation, so that, for the purpose
of comparison we have
\begin{equation}
\Omega^* \propto \frac{\Omega L}{\sqrt{R^{(0)}}} \mbox{  , }
\label{eq:omegastar}
\end{equation}
where the proportionality constant is of order unity. 

In the limit of slow rotation, Ral05
predict a $\Lambda-$effect through the following term:
\begin{equation}
\bar R_{xz} \propto \frac{2 a  }{5}  \sin\gamma \frac{\Omega L}{\sqrt{R^{(0)}}}  
\bar R_{zz}^{(0)}  \mbox{  , }
\end{equation} 
where the proportionality constant is the same as 
in equation (\ref{eq:omegastar}). Meanwhile, our model when written in 
dimensional form and recast in terms of the anisotropy factor $a$  yields
\begin{equation}
\bar R_{xz} = \frac{\left(C_1 - C_6 \frac{a}{a+3}\right) \hat R^{(0)}  }{1-C_6 \hat R^{(0)} (C_1+C_2)} \frac{3(C_1 + C_2)}{3C_1 + C_2} \sin \gamma\, \frac{\Omega L}{\sqrt{R^{(0)}}}  \bar R_{zz}^{(0)}  \mbox{  .}
\end{equation} 
where $ \hat R^{(0)} $ is a dimensionless constant which depends only on the 
model parameters, and is given by equation~(\ref{eq:norotlimit}) with Ro$^{-1}=0$.
In the same slow-rotation limit, the other off-diagonal components of the stress 
tensor are $O({\rm Ro}^{-2})$ or higher order in both our and their models.

Overall, the two formalisms agree on the dependence of the stresses on the rotation rate and 
on latitude, as expected on dimensional and geometrical grounds. In addition, both 
models explicitly demonstrate the importance of the anisotropy of the background non-rotating 
turbulence
in controlling the amplitude of the $\Lambda$-effect. However, the dependence of $\bar R_{xz}$
on the anisotropy factor $a$ superficially appears to be different in the two theories. 
We interpret this in two ways. First, note that the anisotropy factor 
$a$ is a ``free'' parameter in the works of Ral05.
In our model by contrast, there is no freedom in independently 
specifying $a$ since it is a solution of the model once 
the system is specified (e.g. shearing flow, convective flow) and 
depends on the $\{C_i\}$ parameters. In the HRB system for example 
$a = -6C_1  /(3C_1 + C_2) $. 

Secondly, $\bar R_{xz}$ is directly proportional to $a$ in the model of Ral05 while our model 
reveals an additional contribution to the $\Lambda$-effect arising from the background turbulent
heat flux (see equation (\ref{eq:leffect}) for a more explicit expression). 
This contribution is missing from the model of Ral05 which
does not take into account the heat equation. As a result, one may 
superficially conclude that the $\Lambda$-effect could exist even 
for isotropic background turbulent convection. 
In practice, it is difficult to 
conceive of a naturally occurring isotropic turbulent 
system which has a non-zero vertical heat flux, so the
term $\hat F_z^{(0)}$ is in fact also indirectly related 
to the anisotropy of the system, although perhaps not exactly in the same way. 

Finally, we emphasize that in the limit of rapid rotation, neither theory 
is expected to be accurate because of the extreme induced anisotropy of the
rotating turbulent motions. Nevertheless it is interesting to note that the
predicted dependence of the stresses on the rotation rates now no 
longer agree with one another. We find that $\bar R_{yz}$ tends 
to a constant independent of rotation rate while Ral05
find that $\bar R_{yz} \propto$ Ro. For the other off-diagonal 
components $\bar R_{xz}$ and $\bar R_{xy}$ we find a dependence on Ro, 
while they predict a dependence on Ro$^{2}$. 

We conclude this section by emphasizing the success of our closure model in reproducing 
numerical experiments of HRB convection at various aspect ratios and Rayleigh numbers. Furthermore
our model predictions are exactly proportional to those of Ral05 (with a proportionality
constant of order unity) for convection in a slowly rotating system. Hence we 
expect to recover many of the results and successes of these authors in modeling 
differential rotation in stars, albeit with an extended model which self-consistently includes
heat transport in addition to angular momentum transport. In preparation of this future
modeling endeavour, we finally turn to the next natural step of this work, namely the extension of the 
model to the anelastic and fully compressible equations.

\section{The anelastic system and compressible flows}
\label{sec:anelastic}

So far we have worked within the Boussinesq approximation, which is 
applicable only to a shallow layer of fluid whose depth is much less 
than the density scaleheight.  In order to apply our model to stars 
we must first adapt it to the anelastic approximation 
(Ogura \& Phillips 1962; Gough 1969), which is relevant
to subsonic convection in a deep layer.

Here we follow the more standard derivation of the anelastic approximation where
the reference state is taken to be an
adiabatically stratified fluid in hydrostatic equilibrium. The reference density
$\rho_0(\br)$ and temperature $T_0(\br)$ may vary
substantially, while the specific entropy $s_0$ is uniform.  In place
of equations (\ref{boussinesq1})--(\ref{boussinesq3}) we have
\begin{equation}
  \partial_i(\rho_0u_i)=0,
  \label{anelastic1}
\end{equation}
\begin{equation}
  (\partial_t+u_j\partial_j)u_i=-(s-s_0)\partial_iT_0-\partial_i\psi+\cdots,
  \label{anelastic2}
\end{equation}
\begin{equation}
  (\partial_t+u_i\partial_i)(s-s_0)=\cdots,
  \label{anelastic3}
\end{equation}
where the dots represent terms due to viscosity (in the equation of
motion) and thermal conduction (in the thermal energy equation), while
$\psi$ is, again, a modified pressure.  Viscous dissipation can also be included in the thermal energy equation, although it is usually omitted in the Boussinesq approximation.  A derivation of these
equations, omitting diffusive effects, is given in Appendix~B.

The anelastic system is formally very similar to the Boussinesq system
except for the variable density of the reference state.  However, the
entropy perturbation and background temperature gradient respectively play the
roles taken by the temperature perturbation and $\alpha g_i$ in the
Boussinesq approximation.  A very similar analysis to that carried out for
the Boussinesq system leads to equations for $\bar R_{ij}$, $\bar F_i$
and $\bar Q$ of the form
\begin{eqnarray}
  \lefteqn{(\partial_t+\bar u_k\partial_k)\bar R_{ij}+
  \bar R_{ik}\partial_k\bar u_j+\bar R_{jk}\partial_k\bar u_i+
  \bar R_{ij}\partial_k\bar u_k}&\nonumber\\
  &&+\bar F_i\partial_jT_0+\bar F_j\partial_iT_0=\cdots,
\end{eqnarray}
\begin{eqnarray}
  \lefteqn{(\partial_t+\bar u_j\partial_j)\bar F_i+
  \bar R_{ij}\partial_j\bar s+\bar F_j\partial_j\bar u_i+
  \bar F_i\partial_j\bar u_j+\bar Q\partial_iT_0}&\nonumber\\
  &&=\cdots,
\end{eqnarray}
\begin{equation}
  (\partial_t+\bar u_i\partial_i)\bar Q+
  2\bar F_i\partial_i\bar s+\bar Q\partial_i\bar u_i=\cdots,
\end{equation}
where the dots represent terms that require a closure model.  In the
anelastic system the relevant definitions of the Reynolds stress $\bar R_{ij}$,
flux $\bar F_i$ and variance $\bar Q$ are
\begin{equation}
  \bar R_{ij}=\langle\rho_0u_i'u_j'\rangle,\qquad
  \bar F_i=\langle\rho_0u_i's'\rangle,\qquad
  \bar Q=\langle\rho_0s^{\prime2}\rangle.
\end{equation}
Note that $\bar R_{ij}$ now has the correct dimensions for a stress
tensor, and that $\bar F_i$ is really an entropy flux density.  Some
additional linear terms arise in the anelastic system because
$\partial_i\bar u_i\ne0$.

We apply the same closure model as for the Boussinesq system, except
that the relaxation timescale which was proportional to $L/\bar R^{1/2}$ is now 
proportional to $L/(\bar R/\rho_0)^{1/2}$ because of
the redefinition of $\bar R_{ij}$:
\begin{eqnarray}
  \lefteqn{(\partial_t+\bar u_k\partial_k)\bar R_{ij}+
  \bar R_{ik}\partial_k\bar u_j+\bar R_{jk}\partial_k\bar u_i+
  \bar R_{ij}\partial_k\bar u_k}&\nonumber\\
  &&+\bar F_i\partial_jT_0+\bar F_j\partial_iT_0=
  -\frac{C_1}{L}\left({{\bar R}\over{\rho_0}}\right)^{1/2}\bar R_{ij}\nonumber\\
  &&\qquad-\frac{C_2}{L}\left({{\bar R}\over{\rho_0}}\right)^{1/2}
  (\bar R_{ij}-{\textstyle{{1}\over{3}}}\bar R\delta_{ij}),
  \label{rij_anelastic}
\end{eqnarray}
\begin{eqnarray}
  \lefteqn{(\partial_t+\bar u_j\partial_j)\bar F_i+
  \bar R_{ij}\partial_j\bar s+\bar F_j\partial_j\bar u_i+
  \bar F_i\partial_j\bar u_j+\bar Q\partial_iT_0}&\nonumber\\
  &&=-\frac{C_6}{L}\left({{\bar R}\over{\rho_0}}\right)^{1/2}\bar F_i,
  \label{fi_anelastic}
\end{eqnarray}
\begin{equation}
  (\partial_t+\bar u_i\partial_i)\bar Q+
  2\bar F_i\partial_i\bar s+\bar Q\partial_i\bar u_i=
  -\frac{C_7}{L}\left({{\bar R}\over{\rho_0}}\right)^{1/2}\bar Q.
  \label{q_anelastic}
\end{equation}
We do not include any of the terms proportional to $\nu$ or $\kappa$ here 
because we consider the high-Rayleigh number limit in the absence of rigid boundaries only.

The question arises as to how the length-scale $L$ should be
identified for anelastic convection in a deep layer.  It should probably
related to the pressure scaleheight or density scaleheight, as in the stellar mixing-length theory.  
Indeed, numerical simulations of convection in spherical shells with a substantial density variation indicate that the convective cells are much smaller near the outer surface where the scaleheight is small; nevertheless, there may be situations in which convective plumes can span several scaleheights.

Equations~(\ref{rij_anelastic})--(\ref{q_anelastic}) can then be combined with equations for the mean variables in the form
\begin{equation}
  \partial_i(\rho_0\bar u_i)=0,
\end{equation}
\begin{equation}
  \rho_0(\partial_t+\bar u_j\partial_j)\bar u_i=-(\bar s-s_0)\partial_iT_0-
  \rho_0\partial_i\bar\psi-\partial_j\bar R_{ij},
\end{equation}
\begin{equation}
  \rho_0T_0(\partial_t+\bar u_i\partial_i)(\bar s-s_0)=
  \frac{C_1}{L}\left({{\bar R}\over{\rho_0}}\right)^{1/2}{{\bar R}\over{2}}
  -T_0\partial_i\bar F_i.
\end{equation}
In the last equation we have included the turbulent viscous heating.

These equations could be applied to studying convection and meridional circulation in rotating stars.  The solution can be assumed to be axisymmetric and independent of time, although for practical purposes it may be easier to evolve the equations forwards in time until a steady state is reached (if it is) rather than directly seeking such a solution.

In the absence of rotation the problem becomes spherically symmetric, the mean flow disappears, the stress becomes diagonal (although anisotropic) and we obtain the local algebraic system
\begin{equation}
  2\bar F_r\partial_rT_0=-\frac{C_1+C_2}{L}\left({{\bar R}\over{\rho_0}}\right)^{1/2}\bar R_{rr}+\frac{C_2}{3L}\left({{\bar R}\over{\rho_0}}\right)^{1/2}\bar R,
\end{equation}
\begin{equation}
  2\bar F_r\partial_rT_0=-\frac{C_1}{L}\left({{\bar R}\over{\rho_0}}\right)^{1/2}\bar R,
\end{equation}
\begin{equation}
  \bar R_{rr}\partial_r\bar s+\bar Q\partial_rT_0=
  -\frac{C_6}{L}\left({{\bar R}\over{\rho_0}}\right)^{1/2}\bar F_r,
\end{equation}
\begin{equation}
  2\bar F_r\partial_r\bar s=
  -\frac{C_7}{L}\left({{\bar R}\over{\rho_0}}\right)^{1/2}\bar Q.
\end{equation}
The solution is, by direct analogy with equations~(\ref{eq:dimsols1})--(\ref{eq:dimsols2}),
\begin{eqnarray}
  &&\bar R_{rr}=\left({{3C_1+C_2}\over{C_1+C_2}}\right)
  {{\bar R}\over{3}},\nonumber\\
  &&\bar R_{\theta\theta}=\bar R_{\phi\phi}=\left({{C_2}\over{C_1+C_2}}\right)
  {{\bar R}\over{3}},\nonumber\\
  &&\bar F_r=-{{C_1(\bar R/\rho_0)^{3/2}}\over{2L(-N^2)}}\rho_0\partial_r\bar s,\nonumber\\
  &&\bar Q={{C_1\bar R}\over{C_7(-N^2)}}(\partial_r\bar s)^2,
\end{eqnarray}
with
\begin{equation}
  \bar R={{2}\over{C_1C_6}}\left[{{C_1}\over{C_7}}+
  {{3C_1+C_2}\over{3(C_1+C_2)}}\right]\rho_0L^2(-N^2),
\end{equation}
where now
\begin{equation}
  -N^2=(\partial_rT_0)\partial_r\bar s.
\end{equation}
In this situation the entropy gradient $\bar s$ is not known in
advance.  However, to balance the thermal energy equation,
$\partial_i\bar F_i=0$,  which implies that $r^2\bar F_r$ is a
constant, determined by the luminosity generated by the core of the
star.  (This conclusion is modified if the radiative flux or any
sources of energy such as turbulent viscous dissipation make an
important contribution to the thermal energy equation.)  Then the
above equations can be solved algebraically to find $\partial_r \bar s$,
$\bar R$, etc., at each
radius, assuming that a prescription for $L$ is given.  The result is
equivalent to a version of mixing-length theory.

Rotation couples radial and latitudinal transport of heat and momentum
and induces large-scale entropy gradients and mean flows. 
However, if we assume that their effects can be ignored in the overall 
turbulent dynamics controlling the properties of the stresses, then 
the local $\Lambda-$effect is easily recovered as an anelastic version of 
equation (\ref{eq:leffect}). As before, the 
only differences with the Boussinesq case is that (i) the two terms containing
$\hat R^{(0)}$, which have their origin in the eddy turnover time, should be replaced by 
$\hat R^{(0)} / \rho_0$ and (ii) in expressing  (\ref{eq:leffect})  in dimensional form 
(see equation (\ref{eq:nondimrot})), one must
also replace $\tilde N^2$ by $(\partial_rT_0)\partial_r\bar s$ and 
${\rm d} \bar \Theta/{\rm d}z$ by $\rho_0 \partial_r \bar s$, 
as seen above. The resulting expression then directly links the turbulent transport of
angular momentum and of heat to one another. Since heat transport in this model
is very similar to mixing-length theory, our formalism now provides a simple 
framework in which to combine models of stellar structure with models of
internal stellar dynamics. Note that in practice mean flows and especially 
latitudinal entropy gradients could play a role
in the global dynamics of the system. The whole model should therefore be solved 
self-consistently and globally instead of using (\ref{eq:leffect}). This can only be done 
numerically and is deferred to a subsequent paper.

It is also possible to `import' the model of anelastic convection into
the full set of equations governing the motion of a compressible
fluid.  The idea here is that, while the convection might be assumed
to be subsonic and to obey the anelastic approximation, the mean flow
need not obey these constraints.  An example is convection in an
accretion disc, where the accretion flow, although slow, cannot be
treated in the anelastic approximation with a reference density
profile.  Omitting now the bars on all quantities, and neglecting
self-gravitation (although it can easily be restored), we propose a
system of equations consisting of the equation of mass conservation,
\begin{equation}
  \partial_t\rho+\partial_i(\rho u_i)=0,
\end{equation}
the equation of motion,
\begin{equation}
  \rho(\partial_t+u_j\partial_j)u_i=-
  \rho\partial_i\Phi-\partial_ip-\partial_jR_{ij},
\end{equation}
and the thermal energy equation,
\begin{equation}
  \rho T(\partial_t+u_i\partial_i)s=
  \frac{C_1}{L}\left({{R}\over{\rho}}\right)^{1/2}{{R}\over{2}}
  -T\partial_iF_i,
\end{equation}
together with the equations of the closure model,
\begin{eqnarray}
  \lefteqn{(\partial_t+u_k\partial_k)R_{ij}+
  R_{ik}\partial_ku_j+
  R_{jk}\partial_ku_i+R_{ij}\partial_ku_k}&\nonumber\\
  &&+F_i\partial_jT+F_j\partial_iT=
  -\frac{C_1}{L}\left({{R}\over{\rho}}\right)^{1/2}R_{ij}\nonumber\\
  &&\qquad-\frac{C_2}{L}\left({{R}\over{\rho}}\right)^{1/2}
  (R_{ij}-{\textstyle{{1}\over{3}}}R\delta_{ij}),
\end{eqnarray}
\begin{eqnarray}
  \lefteqn{(\partial_t+u_j\partial_j)F_i+R_{ij}\partial_js+
  F_j\partial_ju_i+F_i\partial_ju_j+
  Q\partial_iT}&\nonumber\\
  &&=-\frac{C_6}{L}\left({{R}\over{\rho}}\right)^{1/2}F_i,
\end{eqnarray}
\begin{equation}
  (\partial_t+u_i\partial_i)Q+2F_i\partial_is+
  Q\partial_iu_i=
  -\frac{C_7}{L}\left({{R}\over{\rho}}\right)^{1/2}Q.
\end{equation}
The total energy is then exactly conserved in the form
\begin{eqnarray}
  \lefteqn{\partial_t\left[\rho({\textstyle{{1}\over{2}}}u^2+\Phi+
  e)+{\textstyle{{1}\over{2}}}R\right]}&\nonumber\\
  &&+\partial_i\left[\rho({\textstyle{{1}\over{2}}}u^2+\Phi+h)
  u_i+
  {\textstyle{{1}\over{2}}}Ru_i+R_{ij}u_j+
  TF_i\right]=0,\nonumber\\
\end{eqnarray}
where $e$ and $h$ are the specific internal energy and the specific
enthalpy, respectively, and the gravitational potential $\Phi$ is
assumed to be independent of time.  The existence of this conservation
law implies a certain self-consistency in the equations of the model.
The terms that were added in passing to the compressible model are
required to have the form that they do in order that energy be
conserved.  We note again that $F_i$ is really an entropy
flux density, and that $TF_i$ is the corresponding energy flux density.

The physical content of this model is that the turbulent convecting
fluid behaves similarly to a complex, non-Newtonian material in which
there is a dynamical constitutive equation that relates the stress
tensor to the deformation history of the fluid.  The above equation
for $\partial_tR_{ij}$ (along with those for $\partial_tF_i$ and
$\partial_tQ$) plays this role. 


\section{Conclusions and future prospects}

We have laid out the foundations of a new second-order closure model
for the dynamics of turbulent convection, with future applications 
to stellar convective regions in mind. This model is a direct extension 
of the work of Ogilvie (2003) and GO05, and has similar
properties. The proposed closure has a straightforward physical interpretation, 
and well-understood limitations. 

Comparison with laboratory and numerical experiments reveals
good overall agreement of the model predictions with known properties 
of rotating shear flows (GO05) and high Rayleigh-number
rotating convection (this work). In particular, our model naturally 
reproduces the standard scaling relationships between the Rayleigh and Nusselt
numbers for Rayleigh-B\'enard convection and for Homogeneous Rayleigh-B\'enard 
convection, and contains the well-known $\Lambda$-effect describing 
angular momentum transport in a rotating turbulent fluid. 

When extended to the anelastic (or fully compressible) case, our formalism
can be applied to study convection in stellar interiors. Note that the effects of
Maxwell stresses can also straightforwardly be included following Ogilvie (2003) 
if needed. We show that the model naturally reduces to a version of mixing-length theory when applied 
in a one-dimensional framework. In the presence of rotation
it becomes a powerful tool to study within a single framework the multi-dimensional
balance involving large-scale mean quantities such as the entropy
profile, the meridional circulation and the differential rotation. 
Future work applying our closure model in a spherical shell geometry 
will help understand some of the trends seen in the increasingly 
large number of available observations of stellar differential rotation. 



\section*{Acknowledgements}
The authors thank N. Brummell and C. Doering for stimulating discussions, 
and DuPuits et al. for providing 
electronic tables of their data. This work was supported by 
NSF-AST-0607495 and NSF-CAREER.


\appendix

\section{Realizability}

In this Appendix we show that, provided that the condition
\begin{equation}
  2C_6-C_7-C_1-C_2\ge0
  \label{realizability}
\end{equation}
is satisfied, the evolutionary model used in this paper ensures that the Reynolds stress $\bar R_{ij}$ remains
positive semi-definite and the entropy variance $\bar Q$ remains positive at all points in the flow if they are so initially.  This will ensure that the quantities $\bar R_{ij}$, $\bar
F_i$ and $\bar Q$ predicted by the model can be realized by genuine
velocity and entropy perturbations.  Not least, it will ensure that
the turbulent kinetic energy remains non-negative.  We work here with the anelastic version of the closure model, equations~(\ref{rij_anelastic})--(\ref{q_anelastic}), although a similar argument applies to the Boussinesq system in the high-Rayleigh number limit in which the terms proportional to $\nu$ or $\kappa$ are omitted.

Let $A_i$ be any vector with appropriate dimensions, and consider the
tensor (within the anelastic system)
\begin{equation}
  S_{ij}=\Big\langle\rho_0(u_i'+A_is')(u_j'+A_js')\Big\rangle
\end{equation}
at any point in the flow.  The associated quadratic form is
\begin{equation}
  \sS=S_{ij}X_iX_j=\Big\langle\rho_0\left[(\bu+\bA s')\cdot\bX\right]^2
  \Big\rangle.
\end{equation}
Evidently $\sS\ge0$ for all vectors $X_i$, and therefore $S_{ij}$ must
be a positive semi-definite tensor, for any choice of $A_i$, at every
point in the flow.  Allowing the vector $A_i$ to vary provides us with a family of realizability conditions.

On the other hand, $S_{ij}$ can be expanded as
\begin{equation}
  S_{ij}=\bar R_{ij}+\bar F_iA_j+\bar F_jA_i+\bar QA_iA_j,
\end{equation}
and therefore
\begin{eqnarray}
  \sS&=&\bar R_{ij}X_iX_j+2(\bar{\bF}\cdot\bX)(\bA\cdot\bX)+
  \bar Q(\bA\cdot\bX)^2\nonumber\\
  &=&(\bar R_{ij}-\bar Q^{-1}\bar F_i\bar F_j)X_iX_j+
  \bar Q^{-1}\left[(\bar{\bF}+\bar Q\bA)\cdot\bX\right]^2.\nonumber\\
\end{eqnarray}
Provided that $\bar Q>0$, a necessary and sufficient condition for
$\sS$ to be non-negative for all choices of $A_i$ and $X_i$ is that
the tensor
\begin{equation}
  T_{ij}=\bar R_{ij}-\bar Q^{-1}\bar F_i\bar F_j
\end{equation}
be positive semi-definite.  If this condition is satisfied then $T_{ij}X_iX_j\ge0$ for all $X_i$ and therefore $\sS\ge0$ for all $X_i$ and $A_i$.  On the other hand, if a vector $X_i$ exists such that $T_{ij}X_iX_j<0$, then $\sS<0$ for this choice of $X_i$ if we set $A_i=-\bar F_i/\bar Q$.

We therefore aim to show that, provided that the condition~(\ref{realizability})
is satisfied, the model ensures that $T_{ij}$ remains
positive semi-definite at all points in the flow if, in the initial
state, $T_{ij}$ is positive semi-definite and $\bar Q>0$ at all
points.


We apply a reduction ad absurdum.  If $T_{ij}$ has a negative
eigenvalue at any event, then the quadratic form
\begin{equation}
  \sT=T_{ij}X_iX_j
\end{equation}
is negative at that event, for some choice of the vector $X_i$.
Without loss of generality, let $X_i$ be a differentiable vector field
advected according to the time-reversible equation
\begin{equation}
  {\rm D}X_i-X_j\partial_i\bar u_j=0,
\end{equation}
and such that $\sT<0$ at the event in question.  Here ${\rm
  D}=\partial_t+\bar u_i\partial_i$ is the Lagrangian derivative
following the mean flow.  Retracing the the value of $\sT$ to the
initial state, following the mean flow in reverse, we deduce that
$\sT$ must have passed through zero with ${\rm D}\sT<0$.  However,
using equations (\ref{rij_anelastic})--(\ref{q_anelastic}) we find
\begin{eqnarray}
  \lefteqn{{\rm D}\sT=-\sT\partial_i\bar u_i+
  2\bar Q^{-1}(\bar{\bF}\cdot\bX)X_iT_{ij}\partial_j\bar s}&\nonumber\\
  &&-(C_1+C_2)L^{-1}\left({{\bar R}\over{\rho_0}}\right)^{1/2}\sT+
  C_2L^{-1}\left({{\bar R}\over{\rho_0}}\right)^{1/2}{{\bar R}\over{3}}X^2
  \nonumber\\
  &&+(2C_6-C_7-C_1-C_2)L^{-1}\left({{\bar R}\over{\rho_0}}\right)^{1/2}
  \bar Q^{-1}(\bar{\bF}\cdot\bX)^2.\nonumber\\
\end{eqnarray}
When $\sT=0$, $X_i$ is a null eigenvector of $T_{ij}$ and therefore
${\rm D}\sT\ge0$, with equality only when there is no turbulence.
Therefore $T_{ij}$ cannot in fact develop a negative eigenvalue.


\section{Numerical algorithm for Rayleigh--B\'enard convection}

The equations governing Rayleigh-B\'enard-convection, (9-11), can be written 
in the standard non-dimensional form 
\begin{eqnarray}
\partial_t {\bf {\hat u}}
- Pr \nabla^2 {\bf {\hat u}}
& = &
- \nabla \hat{\psi}
+ {\rm Pr}\,{\rm  Ra}\, \hat{T} {\bf e}_z 
- (\nabla \times {\bf {\hat u}}) \times {\bf {\hat u}} \; ,\label{eq:momentum} \\
\partial_t \hat{T}
- \nabla^2 \hat{T}
& = &
-\nabla \cdot ({\bf {\hat u}}  \hat{T}) \; ,\label{eq:Temperature}\\
\nabla \cdot {\bf {\hat u}}  &=&  0 \; ,
\label{eq:continuity}
\end{eqnarray}
where ${\bf e}_z$ is the vertical unit vector and 
non-dimensional quantities are marked by a hat. Here 
the layer height $h$ is used as the unit length, 
$h^2/\kappa$ is the unit time and the imposed
temperature difference between the two plates $\Delta T$  
serves as the temperature scale. 

We decompose the velocity field ${\bf \hat u} = (\hat u_x,\hat u_y,\hat u_z)$ into toroidal, poloidal and mean parts, 
\begin{equation}
{\bf {\hat u}} =  \nabla \times (\hat{e} \,  {\bf e}_z) +  \nabla \times  \nabla \times (\hat{f} \,  {\bf e}_z) +
\langle \hat{u}_{x} \rangle_{h}  {\bf e}_x + \langle \hat{u}_{y} \rangle_{h}  {\bf e}_y \mbox{  ,}
\label{eq:tor_pol_mean_deomp}
\end{equation}
where  $\langle ... \rangle_{h}$ denotes a horizontal average.
Note that (\ref{eq:tor_pol_mean_deomp}) automatically satisfies (\ref{eq:continuity}). Equations for the 
scalar functions $\hat{e}$ and $\hat{f}$ and for the horizontally averaged velocities
$ \langle \hat u_{x} \rangle_{h}$ and $ \langle \hat u_{y} \rangle_{h}$ can be derived 
by applying the operators ${\bf e}_z \cdot {\bf  \nabla} \times$ and 
${\bf e}_z \cdot  {\bf  \nabla} \times {\bf  \nabla} \times$ to the momentum equation 
(\ref{eq:momentum}) and by averaging this equation horizontally, leading
to 
\begin{eqnarray}
(\partial_t - {\rm Pr} \nabla^2) \nabla^2_H \hat{e}
&=&  {\bf e}_z \cdot  \nabla \times {\bf N}  \mbox{  , }
\label{eqn:toroidal}\\
(\partial_t - {\rm  Pr} \nabla^2) \nabla^2_H \nabla^2 \hat{f}
&=& -{\bf e}_z \cdot  \nabla \times  \nabla \times {\bf N} \nonumber \\  
&-& {\rm Pr}\, {\rm Ra} \, \nabla^2_H \hat T  \mbox{  , }
\label{eqn:poloidal}\\
(\partial_{t} - {\rm  Pr} \; \partial_{z}^{2}) \langle \hat{u}_{x} \rangle_{h}
&=& - {\bf e}_x \cdot \langle {\bf N} \rangle_{h} \mbox{  , }
\label{eqn:mean_ux}\\
(\partial_{t} -{\rm  Pr} \; \partial_{z}^{2}) \langle \hat{u}_{y} \rangle_{h}
&=& - {\bf e}_y \cdot \langle {\bf N}  \rangle_{h}\mbox{  , }
\label{eqn:mean_uy}
\end{eqnarray}
where $\nabla^2_H = \partial_x^2+\partial_y^2$ is the horizontal Laplacian and where
the vector quantity ${\bf N}$ is defined as
\begin{equation}
{\bf N} = (\nabla \times {{\bf {\hat u}}}) \times {\bf {\hat u}} \mbox{  . }
\end{equation} 
Either stress-free or no-slip boundary conditions may be applied for ${\bf {\hat u}}$, which translates into the boundary conditions 
\begin{eqnarray}
&& \partial_z \hat{e} = \hat{f} = \partial_z^2 \hat{f} = \partial_z \langle \hat{u}_x \rangle = \partial_z \langle \hat{u}_y \rangle = 0 \quad  \mbox{(stress free)} \; \nonumber \\
&& \hat{e} = \partial_z \hat{f} = \partial_z^2 \hat{f} = \langle \hat{u}_x \rangle = \langle \hat{u}_y \rangle = 0 \quad \mbox{(no slip)}\; 
\end{eqnarray}
at $ \: z=0$ and $z=1$ for the toroidal and poloidal scalars $\hat{e}$ and $\hat{f}$ and for 
the mean velocities $\langle \hat{u}_x \rangle$ and $\langle \hat{u}_y \rangle$.

A pseudo-spectral algorithm is used to solve the governing equations  
in the formulation (\ref{eq:Temperature},\ref{eqn:toroidal}-\ref{eqn:mean_uy}).
The fields $\hat{e},\hat{f}$ and $\hat{T}$ are expanded as Fourier series in the horizontal
direction. In the vertical direction, a Chebyshev expansion on a Gauss-Lobatto 
grid is used for all unknowns. Fast transform algorithms can then be applied to 
switch between physical and transform space.
A semi-implicit time-stepping scheme is employed for the temporal discretization, 
where all linear terms are 
treated implicitly by a second order Backward-Differencing (BDF2) scheme, while a second order 
Adams-Bashforth (AB2) scheme is applied to the nonlinear terms. 

Most of the computation is carried out in spectral space, although the 
nonlinear terms are evaluated in physical space. The usual $3/2$-rule is
applied to avoid aliasing errors in the horizontal directions, whereas no
de-aliasing procedure is employed along the vertical coordinate. 
We use the Chebyshev tau method (Peyret, 2002) to solve the ODEs
arising from the implicit part of the time-stepping scheme. 
This method has the advantage of yielding linear systems which can be manipulated 
into sparse form, thus keeping the memory requirements at a 
manageable level. The code is parallelized using transpose-based 
parallel FFTs, see Stellmach \& Hansen (2008) for details. 

\section{Numerical algorithm for Homogeneous Rayleigh--B\'enard convection}

A spectral algorithm using Fourier expansions in all three spatial directions 
is used to solve the governing equations (49) in the homogeneous Rayleigh-B\'enard case. 
The primitive variables ${\bf u'}$,$T'$,$\psi'$ 
are used, with the pressure perturbation $\psi'$ being calculated in the same way as in the 
classical Patterson-Orzag Algorithm (Canuto et al. 2007) which is widely used in simulations 
of homogeneous, isotropic turbulence. Non-linear products are evaluated on a grid in physical 
space and aliasing errors are avoided by using the $3/2$-rule. The equations are advanced in
time by a semi-implicit multistep method in which all diffusive terms are treated implicitly by
a third order Backward-Differencing (BDF3) algorithm, while a third-order Adams-Bashforth 
(AB3) scheme is applied to the non-linear terms. This time stepping 
method offers a relatively large stability domain that includes a part of the imaginary axis
at a comparatively low cost (Peyret, 2002). As a starting scheme, we use a second-order Runge-Kutta 
method. A parallelization approach similar to the one employed in our Rayleigh-B\'enard 
convection code is used, see Stellmach \& Hansen (2008) for details.

\section{Derivation of the anelastic system}

The equations governing the motion of an ideal, compressible fluid are
the equation of mass conservation,
\begin{equation}
  \partial_t\rho+\partial_i(\rho u_i)=0,
\end{equation}
the equation of motion,
\begin{equation}
  \rho(\partial_t+u_j\partial_j)u_i=-\rho\partial_i\Phi-\partial_ip,
\end{equation}
the thermal energy equation,
\begin{equation}
  \rho T(\partial_t+u_i\partial_i)s=0,
\end{equation}
and Poisson's equation,
\begin{equation}
  \partial_{ii}\Phi=4\pi G\rho.
\end{equation}
By introducing the specific enthalpy $h$, the equation of motion can
be rewritten in the form
\begin{equation}
  (\partial_t+u_j\partial_j)u_i=-\partial_i(\Phi+h)+T\partial_is.
\end{equation}
We adopt a system of units in which the pressure scale-height and the
sound speed are of order unity.  We introduce a small parameter
$\epsilon$ such that the (imaginary) Brunt--V\"ais\"al\"a frequency is
$O(\epsilon)$ when expressed in these units.  We then pose the
asymptotic expansions
\begin{eqnarray}
  \rho&=&\rho_0(\br)+\epsilon^2\rho_2(\br,\tau)+O(\epsilon^4),\nonumber\\
  \bu&=&\epsilon\bu_1(\br,\tau)+O(\epsilon^3),\nonumber\\
  \Phi&=&\Phi_0(\br)+\epsilon^2\Phi_2(\br,\tau)+O(\epsilon^4),\nonumber\\
  h&=&h_0(\br)+\epsilon^2h_2(\br,\tau)+O(\epsilon^4),\nonumber\\
  T&=&T_0(\br)+\epsilon^2T_2(\br,\tau)+O(\epsilon^4),\nonumber\\
  s&=&s_0+\epsilon^2s_2(\br,\tau)+O(\epsilon^4),
\end{eqnarray}
where $\tau=\epsilon t$ is a slow time variable.  Note that the
reference state is adiabatically stratified, and therefore $s_0$ is
independent of $\br$.  The equation of motion at leading order [$O(1)$] requires
hydrostatic equilibrium in the reference state,
\begin{equation}
  0=-\partial_i(\Phi_0+h_0),
\end{equation}
while Poisson's equation at leading order [$O(1)$] is
\begin{equation}
  \partial_{ii}\Phi_0=4\pi G\rho_0.
\end{equation}
At $O(\epsilon^2)$ the equation of motion gives
\begin{equation}
  (\partial_\tau+u_{1j}\partial_j)u_{1i}=-\partial_i(\Phi_2+h_2)+
  T_0\partial_is_2.
\end{equation}
This can also be written in the form
\begin{equation}
  (\partial_\tau+u_{1j}\partial_j)u_{1i}=-s_2\partial_iT_0-\partial_i\psi,
\end{equation}
where $\psi=\Phi_2+h_2-T_0s_2$ is a modified pressure variable.  The equation of mass conservation at
leading order [$O(\epsilon)$] is
\begin{equation}
  \partial_i(\rho_0u_{1i})=0,
\end{equation}
and the thermal energy equation at leading order [$O(\epsilon^3)$] is
\begin{equation}
  \rho_0T_0(\partial_\tau+u_{1i}\partial_i)s_2=0.
\end{equation}
Poisson's equation at $O(\epsilon^2)$ is not required, and the
departures from the reference state are not affected by self-gravitation at leading order.  When the
asymptotic scalings are removed, and allowance is made for diffusive
effects, equations (\ref{anelastic1})--(\ref{anelastic3}) are
obtained.

\label{lastpage}

\end{document}